\documentclass[aps,prb,twocolumn,10pt,a4paper,footinbib,floatfix]{revtex4}
\usepackage{graphicx,amsmath}
\usepackage{amssymb}
\usepackage{amsthm}
\usepackage{hyperref}
\usepackage{doi}
\allowdisplaybreaks[4]
\usepackage{tikz}
\usepackage[caption=false]{subfig}
\usepackage{lipsum}

\begin{document}
\title{Study of the phase diagram of the Kitaev-Hubbard chain}
\author{Iman Mahyaeh}
\author{Eddy Ardonne}
\affiliation{Department of Physics, Stockholm University, SE-106 91 Stockholm, Sweden}
\date{\today}

\begin{abstract}
We present a detailed study of the phase diagram of the Kitaev-Hubbard chain, that is the Kitaev chain in the presence of a nearest-neighbour density-density interaction, using both analytical techniques as well as DMRG. In the case of a moderate attractive interaction, the model has the same phases as the non-interacting chain, a trivial and a topological phase. For repulsive interactions, the phase diagram is more interesting. Apart from the previously observed topological, incommensurate and charge density wave phases, we identify the `excited state charge density wave' phase. In this phase, the ground state resembles an excited state of an ordinary charge density phase, but is lower in energy due to the frustrated nature of the model. We find that the dynamical critical exponent takes the value $z\simeq 1.8$. Interestingly, this phase only appears for even system sizes, and is sensitive to the chemical potential on the edges of the chain. For the topological phase, we present an argument that excludes the presence of a strong zero mode for a large part of the topological phase. For the remaining region, we study the time dependence of the edge magnetization (using the bosonic incarnation of the model). These results further expand the region where a strong zero mode does not occur. 
\end{abstract}
\maketitle

\section{Introduction}
\label{introduction}
Topological phases of non-interacting fermions have been studied extensively and the full classification of the possible phases has been found\cite{class08,class10,class-RMP,kitaev09}. The classification scheme is based on time-reversal symmetry, particle-hole symmetry and chiral symmetry and which phases are possible depends on the spatial dimension. To a given class and spatial dimension one can attribute a group, i.e. $\mathbb{Z}_2$ or $\mathbb{Z}$, that describes the possible gapped topological phases in that class. As an example the class $\mathbf{BDI}$ has all three symmetries (squaring to one) and in one dimension its gapped phases are labeled by the group $\mathbb{Z}$. The prototypical model belonging to this class is the Kitaev chain (with real couplings)\cite{kitaev01}. In the topological phase of the Kitaev chain, the model hosts Majorana zero modes on the edges of the chain. This zero mode results in a fully doubly degenerate many-body spectrum.

For interacting fermions there is no universal classification scheme. Nonetheless for the class $\mathbf{BDI}$ it was shown that in the presence of interactions the different topological classes correspond to the elements of the group $\mathbb{Z}_8$\cite{fk10,fk11}. One should note that this classification only concerns gapped phases and the phase diagram of a general model has gapless points or regions as well. Therefore it is interesting not only to wonder about how the interaction affects a topological phase\cite{alicea-fisher11} but also how gapless phases emerge due to interactions. 

The main aim of this paper is studying the phase diagram of the Kitaev chain in the presence of a density-density interaction, see Refs.~\onlinecite{selke,feo06,feo07,nagy,sela11,allen} for earlier results for this model. We refer to this model as the Kitaev-Hubbard chain. We study both the attractive and repulsive regime. In addition to the trivial and topological phases which are inherited from the Kitaev chain, the model has an incommensurate phase, a charge density wave phase and a new phase which is sensitive to the system size and the chemical potential at the boundaries of the chain. Moreover we study the possibility of having a full doubly degenerate many-body spectrum in the topological phase of this interacting model.  

The paper is organized as follows. In Sec.~\ref{sec-model} we briefly review both the classical and quantum axial next-nearest-neighbour Ising (ANNNI) model and present the model of interest for this paper, namely the Kitaev-Hubbard chain. We show that the quantum ANNNI model and the Kitaev-Hubbard chain are dual to each other and present the main result, that is the phase diagram of this model. We continue by presenting our analytical results based on bosonization and the numerical results based on density matrix renormalization group (DMRG)\cite{white92,scholl05} for an attractive interaction in Sec.~\ref{sec-attractive}. In Sec.~\ref{sec-repulsive} we move on to the repulsive regime, for which there is an incommensurate phase and a new phase which was missed previously. Moreover, we will discuss the properties of the highest excited states as well as the finite temperature features of the model in Sec.~\ref{sec-excited_and_finite_T}. We conclude the paper in Sec.~\ref{sec-discussion}. 

\section{The Model and the phase diagram}
\label{sec-model}
To introduce the model, we set the scene by reminding the reader of the 
classical Ising model, which is the corner stone of our understanding of phase transitions\cite{onsager,baxter,kardar,goldenfeld}.

To define this model consider a square lattice where on each site, with coordinate $(i,j)$, on has a classical spin, $s_{i,j}=\pm 1$. For a configuration of spins, $\{s\}$ the energy is 
\begin{equation}
E[\{s\}] = -\sum_{i,j} \left(\tilde{J}_1 s_{i,j}s_{i+1,j} + \tilde{J}_{\perp}s_{i,j}s_{i,j+1}\right) \ .
\end{equation}
We assume that $\tilde{J}_1, \tilde{J}_{\perp}>0$. The classical Ising model in two-dimensions was initially solved by Onsager\cite{onsager}, later other solutions were found, see \onlinecite{baxter}. The model has two phases. For temperatures $T$ less than the critical temperature, $T_c$, it is in the ferromagnetic phase and for $T>T_c$ it is in the disordered phase. 

As a generalization one can consider the ANNNI model where one adds an extra interaction along one axis, which could favour ferromagnetic or anti-ferromagnetic alignment\cite{sf80,selke81,selke}. The energy for a given configuration of spins of the ANNNI model
is
\begin{align}
E[\{s\}] = & -\sum_{i,j} \left( \tilde{J}_1 s_{i,j}s_{i+1,j}+\tilde{J}_2 s_{i,j}s_{i+2,j} \right) \nonumber\\
& - \sum_{i,j} \tilde{J}_{\perp} s_{i,j}s_{i,j+1} \ .
\end{align}

We follow the usual notation and define $\tilde{\kappa}=-\frac{\tilde{J}_2}{\tilde{J}_1}$. For $\tilde{\kappa} <0$ there is no frustration, and the axial term only modifies the critical temperature. For $\tilde{\kappa}> 0$, however, the situation is more subtle. For small and positive $\tilde{\kappa}$ the model has two phases, a ferromagnetic and paramagnetic phase. The point $\tilde{\kappa}=1/2$ is a multi-critical point, where four phases meet, namely the ferromagnet, paramagnet, a floating phase and the so-called `anti-phase'\cite{selke}. In the anti-phase, the axial interaction dominates and there are four spin configurations that minimize the energy. For the floating phase however, all interactions are important. Whether the floating phase continues to infinite $\tilde{\kappa}$ is an open question, which we study in this paper by analyzing the corresponding quantum model.

The quantum counterpart of the classical Ising model is the transverse field Ising model (TFIM)\cite{epw70,pfeuty70,fs78}. This model is defined on a chain. On each site there is a spin-$1/2$ degree of freedom and the Hamiltonian for a chain of size $L$ can be written as
\begin{equation}
H=-J_1 \sum_{j=1}^{L-1} \tau^z_j\tau^z_{j+1} + B \sum_{j=1}^{L}\tau^x_j \ , 
\end{equation}
where $\tau^{\alpha}$ are Pauli matrices. One can relate the parameters $J_1$ and $B$ in terms of coupling constants of the classical model and the temperature\cite{fs78}. The TFIM shows a quantum phase transition at $T=0$. For $|B|<|J_1|$ the model is in the ferromagnetic phase while for $|B|>|J_1|$ it is in the disordered phase. The classical Ising model in two dimensions and the one-dimensional TFIM are in the same universality class.

The quantum ANNNI model was also studied\cite{feo06,feo07,nagy,allen}. The Hamiltonian reads\cite{selke}
\begin{equation}
\label{eq-qannni}
H=-J_1 \sum_{j=1}^{L-1} \tau^z_j\tau^z_{j+1} -J_2 \sum_{j=1}^{L-2} \tau^z_j\tau^z_{j+2} + B \sum_{j=1}^{L}\tau^x_j \ . 
\end{equation}
In analogy with the classical ANNNI model, we define $\kappa=-\frac{J_2}{J_1}$. For $\kappa < \frac{1}{2}$ the model has only ferromagnetic and paramagnetic phases. In this regime there is a second order phase transition between the ordered and the disordered phase at $B_c(\kappa)$. The point $\kappa=\frac{1}{2}$ is a multi-critical point from which the floating phase emerges. For very large $\kappa$ the dominant term is the next-nearest-neighbour term in Eq.~\eqref{eq-qannni} which gives rise to four ground states. Previous studies found that the floating phase survives till $\kappa\simeq 5$\cite{feo07,nagy}. We will present our results for infinite $\kappa$ by studying the dual model.

To study large frustration, $\kappa \gg 1$, we employ the duality map as follows:
\begin{align}
\tau^z_j\tau^z_{j+1} &\rightarrow \sigma^z_j \ , \nonumber \\
\tau^x_j & \rightarrow \sigma^z_j\sigma^z_{j+1} \ . 
\end{align}
We further perform an on-site rotation,
\begin{equation}
\sigma^x_j \rightarrow \sigma^z_j \ , \hspace{0.5cm}\sigma^z_j \rightarrow (-1)^j\sigma^x_j \ .
\end{equation}
Using these two transformations one can transform the Hamiltonian for the quantum ANNNI model, Eq.~\eqref{eq-qannni}, to an interacting quantum Ising model,
\begin{equation}
H=-B \sum_{j=1}^{L-1} \sigma^x_j\sigma^x_{j+1} - J_1 \sum_{j=1}^{L} \sigma^z_j +J_2 \sum_{j=1}^{L-1} \sigma^z_j\sigma^z_{j+1} \ .
\end{equation}

Note that the sign of $B$ and $J_1$ are not important and we will assume that both are positive. The sign of $J_2$, however, is crucial. Hence we use $B$ as our energy scale and study the following Hamiltonian, 
\begin{equation}
\label{eq-q-int-ising}
H(h,U)=-\sum_{j=1}^{L-1} \sigma^x_j\sigma^x_{j+1} - h \sum_{j=1}^{L} \sigma^z_j + U\sum_{j=1}^{L-1} \sigma^z_j\sigma^z_{j+1} \ ,
\end{equation}
in which,
\begin{equation}
\label{eq-parameters_dualirty}
h= \frac{J_1}{B} \ , \hspace{0.5cm}U= \kappa \frac{J_1}{B}=\kappa h \ .
\end{equation}
By performing a Jordan-Wigner (JW) transformation\cite{jw28}, one arrives at the
Kitaev-Hubbard chain,
\begin{align}
\label{eq-fermionic-incarnation}
H(h,U)&= - \sum_{j=1}^{L-1}\left( c^{\dagger}_j - c_j\right) \left( c^{\dagger}_{j+1} + c_{j+1}\right) \nonumber \\
& - h \sum_{j=1}^{L} \left( 1-2 c^{\dagger}_j c_j \right) \nonumber \\
& + U \sum_{j=1}^{L-1} \left( 1-2 c^{\dagger}_j c_j \right) \left( 1-2 c^{\dagger}_{j+1} c_{j+1} \right) \ .
\end{align}

For $U=0$, the model reduces to the TFIM (or the Kitaev chain\cite{kitaev01}) which is exactly solvable\cite{pfeuty70,lsm61}. In the bosonic representation the two phases are the ordered ($h < 1$) and the disordered ($h >1$) phases which are converted to the topological and the trivial phases respectively in the fermionic incarnation. The phase transition between these phases is described by the $c=\frac{1}{2}$ Ising CFT.

In the topological phase the model hosts Majorana zero modes which live on the edges and has zero energy up to exponentially small corrections in the system size. An interesting question is whether the topological phase is stable in the presence of interactions\cite{alt11,alicea-fisher11}. The $U$ term in Eq.~\eqref{eq-q-int-ising} gives rise to a density-density interaction, but renders the model non-integrable. 

Another special case is $h=0$ for which the model is called the $XY$ model\cite{lsm61}. This model is  gapped, except when $U=\pm 1$, which are critical points with central charge $c=1$. The gapped phases are ordered phases in the bosonic representation and topological in the fermionic incarnation. 

It is worth to mention that the model in Eq.~\eqref{eq-q-int-ising} is actually relevant for an array of superconducting islands\cite{hs12}. Moreover for a specific set of couplings known as the Peschel-Emery (PE) line, the ground state is exactly doubly degenerate\cite{pe81}. This line lies in the topological phase and host weak zero modes\cite{katsura15,ma18}, see Sec.~\ref{sec-excited_and_finite_T} for more detail on this.

\begin{figure}[h]
\includegraphics[width=\columnwidth]{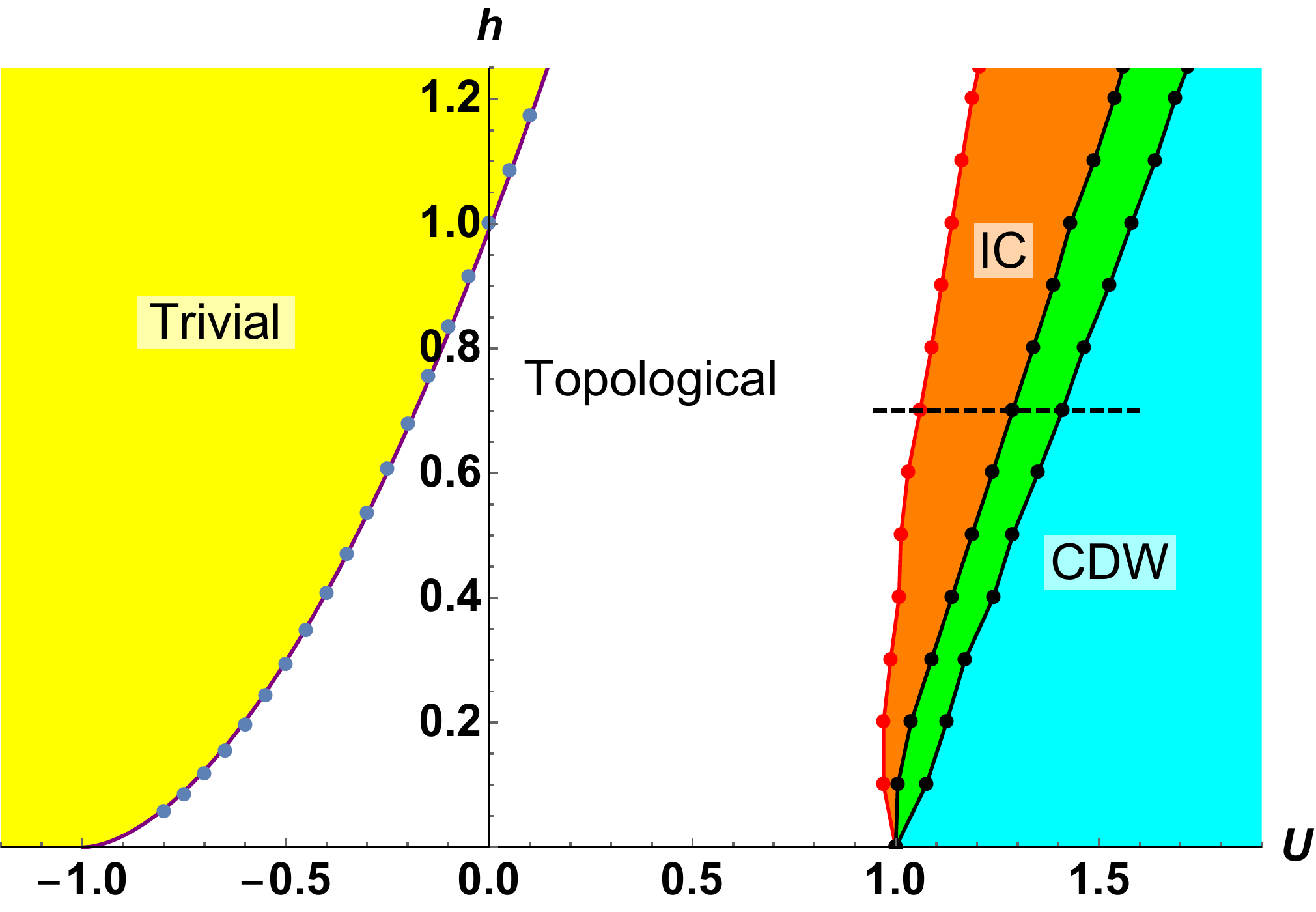}
\caption{(Color online) The Phase diagram for the model in Eq.~\eqref{eq-q-int-ising}. The green region, between the IC and CDW phase, is the esCDW phase. } 
\label{fig-pd}
\end{figure}

We obtained the phase diagram of the model in Eq.~\eqref{eq-q-int-ising} as presented in Fig.~\ref{fig-pd}.
Parts of this phase diagram were schematically drawn previously based on the known results for the quantum ANNNI model\cite{feo06,feo07,sela11,alt11,hs12}. We believe, however, that in the previous studies one phase (the green region in the plot) has been missed. This phase will be described in Sec.~\ref{sec:new-phase}.

\section{The Attractive Interaction case}
\label{sec-attractive}

For attractive interactions, $U<0$, there is no frustration, $\kappa < 0$, and one expects a direct transition from the topological phase (ordered phase) to the trivial phase (disordered phase), as is the case for $U=0$. To study the $U<0$ case, one can start from the Ising critical point, namely $h=1$ and $U=0$, and perform perturbative calculations to find the location of the phase transition\cite{gergs}. We will, however, start from another critical point, namely the $XY$ critical point with $h=0$ and $U=-1$. Close to this point we can use bosonization. To do so it is convenient to transform the Hamiltonian in Eq.~\eqref{eq-q-int-ising}. We first do an on-site rotation,  
\begin{equation}
\sigma^x_j \rightarrow (-1)^j\sigma^y_j  \ , \hspace{0.5cm}\sigma^z_j \rightarrow (-1)^j\sigma^x_j  \ , 
\end{equation}
which changes the Hamiltonian to
\begin{align}
\label{eq-hamil-for Bos}
H(h,U) &=\sum_j  \left( \sigma^x_j\sigma^x_{j+1} + \sigma^y_j\sigma^y_{j+1} \right)  \nonumber \\ 
& + \sum_j \left( - \delta U \sigma^x_j\sigma^x_{j+1} - h (-1)^j \sigma^x_j \right) \ ,
\end{align}
in which we defined $\delta U = U+1$ and dropped the lower and upper bounds of the sums, since we will be working in the continuum limit. We assume that $\delta U, h \ll 1$.  

We can use the JW transformation, write the Hamiltonian in terms of spinless fermions, $\psi_j$, and perform a Fourier transformation to work in momentum space. Doing so, the first two terms, namely the $XY$ model, give rise to a gapless $\cos(k)$ band, which can be linearized around $k=\pm\frac{\pi}{2}$, to get the continuum model,
\begin{equation}
\label{eq-bos-psi_x}
\psi_j = \sqrt{a} \left[ e^{i \frac{\pi}{2} j } \psi_+(x) + e^{-i \frac{\pi}{2} j } \psi_-(x)\right] \ .
\end{equation}
In the last equation $a$ is the lattice spacing and $x=ja$ will be the continuous spatial coordinate. 

We first consider the $\delta U$ term in Eq.~\eqref{eq-hamil-for Bos}. Plugging Eq.~\eqref{eq-bos-psi_x} into this term we get a term proportional to
\begin{equation}
\int \left( \psi^\dagger_+(x) \psi^\dagger_-(x)  + \psi_+(x) \psi_-(x) \right) dx \ .
\end{equation}

To further simplify the result, we use the dual fields $\phi(x)$ and $\theta(x)$, which obey the commutator $[ \phi(x),\theta(y) ] = \mathit{i} \Theta(y-x)$, where $\Theta(x)$ is the Heaviside step function. Therefore, $\partial_x \theta(x)$ is the conjugate momentum of the field $\phi(x)$. We employ the bosonization dictionary\cite{book:shankar},
\begin{equation}
\psi_{\pm}(x) = \frac{1}{\sqrt{2 \pi \alpha}} e^{i \sqrt{\pi} \left(  \pm \phi(x) - \theta(x)  \right) } \ , 
\end{equation}
in which $\alpha$ is a cut-off in momentum space. Hence we can write the $\delta U$ term in terms of the dual fields, which is proportional to
\begin{equation}
\int \cos(\sqrt{4 \pi} \theta) dx  \ .
\end{equation}

\begin{figure}[h]
\includegraphics[width=\columnwidth]{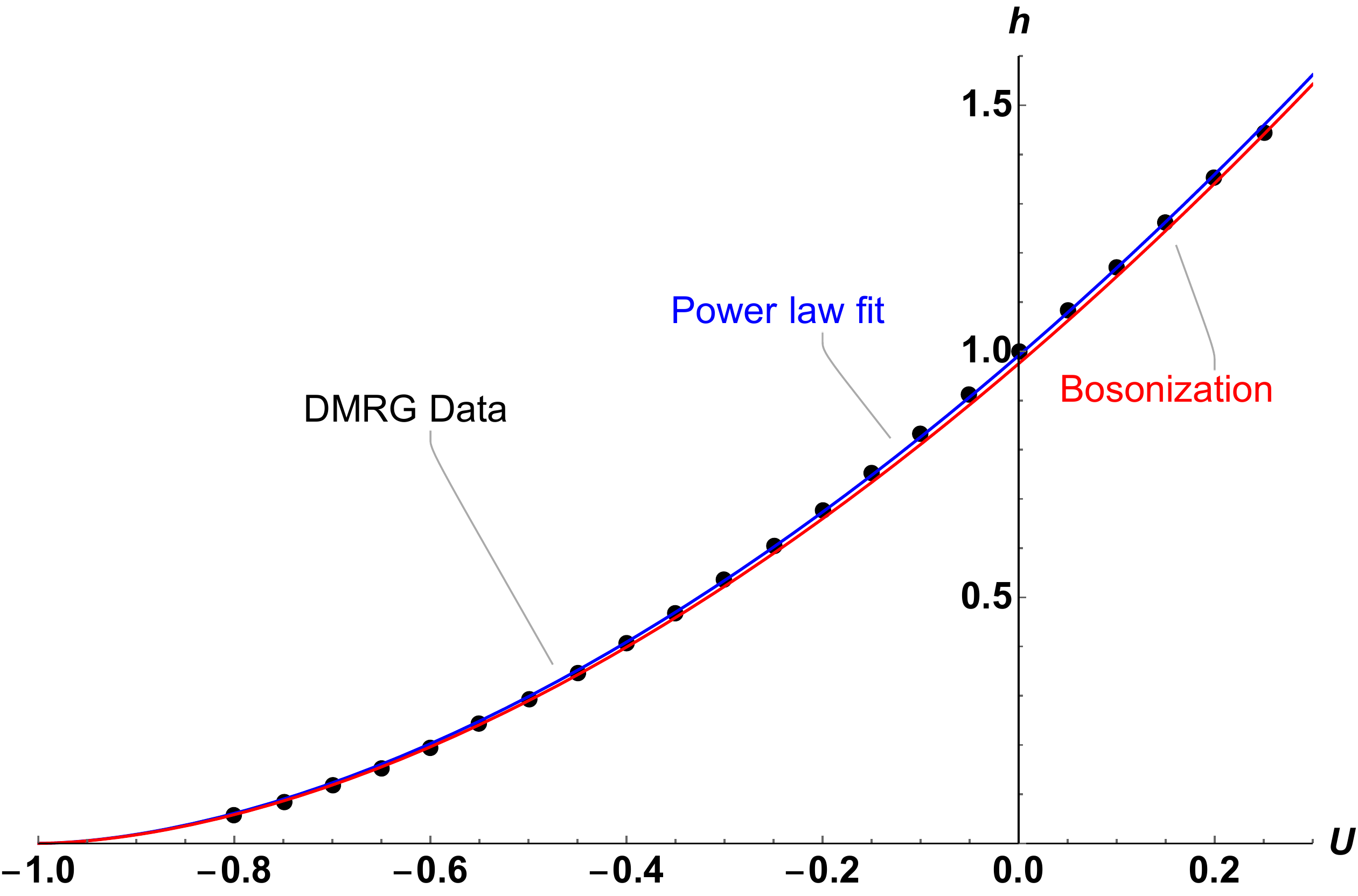}
\caption{(Color online) The phase boundary between the trivial band insulator and the topological phase. The blue line is $h_c(U)=0.99 (U+1)^{1.73}$. The red line is $h_c(U)=0.97 (U+1)^{\frac{7}{4}}$. } 
\label{fig-bos}
\end{figure}

The $h$ term in Eq.~\eqref{eq-hamil-for Bos} can be treated in the same way. Using the bosonization dictionary and by dropping the fast oscillatory terms which are proportional to $(-1)^j$, one can show that
\begin{equation}
(-1)^j \sigma^-_j \rightarrow e^{- i \sqrt{\pi}  \theta(x)} \ .
\end{equation}
Therefore the $h$ term reduces to
\begin{equation}
\int \cos(\sqrt{\pi} \theta) dx  \ .
\end{equation}
The full Bosonized Hamiltonian then reads
\begin{align}
\label{eq-bosonized-hamil}
H(h, 1+ \delta U) & = H_0 + c_1 \delta U \int \cos(\sqrt{4 \pi} \theta) dx   \nonumber \\
&+ c_2 h\int \cos(\sqrt{\pi} \theta) dx \ ,
\end{align}
in which $H_0$ is the free bosonic Hamiltonian and $c_1$ and $c_2$ are some constants which depend on $a$ and $\alpha$. Since $H_0$ is quadratic, we can calculate the renormalization group flow of the couplings, $\delta$ and $h$, up to first order
\begin{eqnarray}
\frac{d \delta U}{dl} = \delta U \ , \\
\frac{d h}{dl} = \frac{7}{4} h \ . 
\end{eqnarray}
We should mention that we assumed that we can drop the renormalization of $K$, one of the Luttinger parameters. These two equations tell us that close to the $XY$ point, $h=0$ and $U=-1$, the curve separating the topological and the trivial phases is given by
\begin{equation}
\label{eq-result-bosonization}
h \sim (U+1)^{\frac{7}{4}} \ . 
\end{equation}
To check this analytical result we performed a numerical study and compared to the results shown in Fig.~\ref{fig-bos}. To locate the phase transition we ran DMRG using the ALPS libraries\cite{alps07,alps11,alps14}. This DMRG study also serves to get acquainted with the performance of the algorithm for the model at hand. 

We should note that the second term in Eq.~\eqref{eq-fermionic-incarnation} (the magnetic field in the bosonic incarnation in Eq.~\eqref{eq-q-int-ising}) has been implemented as follows
\begin{equation}
\label{eq-chemical-bonds}
\sum _{j=1}^{L-1} - \frac{h}{2} \left[  \left( 1-2 c^{\dagger}_j c_j \right) + \left( 1-2 c^{\dagger}_{j+1} c_{j+1} \right)\right] \ .
\end{equation}
Therefore in the bosonic incarnation the strength of the magnetic field in the bulk is $h$ while it is half of this strength on the first and the last sites, $h_1=h_{L}= \frac{h}{2}$. 

We used finite size scaling (FSS) to pinpoint the transition. Within this framework, one finds the ground state and the first excited state and use the following scaling ansatz for the energy difference between them, $\delta$, close to the critical point \cite{fb72},
\begin{equation}
\delta(L,h)=L^{-z} \mathcal{F} \left(L^{-\nu} \left( h-h_c \right) \right) \ ,
\label{eq:scaling-form}
\end{equation}
where $L$ is the system size, $z$ is the dynamical critical exponent, $\nu$ is the correlation length exponent and $\mathcal{F}$ is a scaling function. For the TFIM it is known that $z=\nu=1$ \cite{sachdevbook}.

\begin{figure}[h]
\includegraphics[width=\columnwidth]{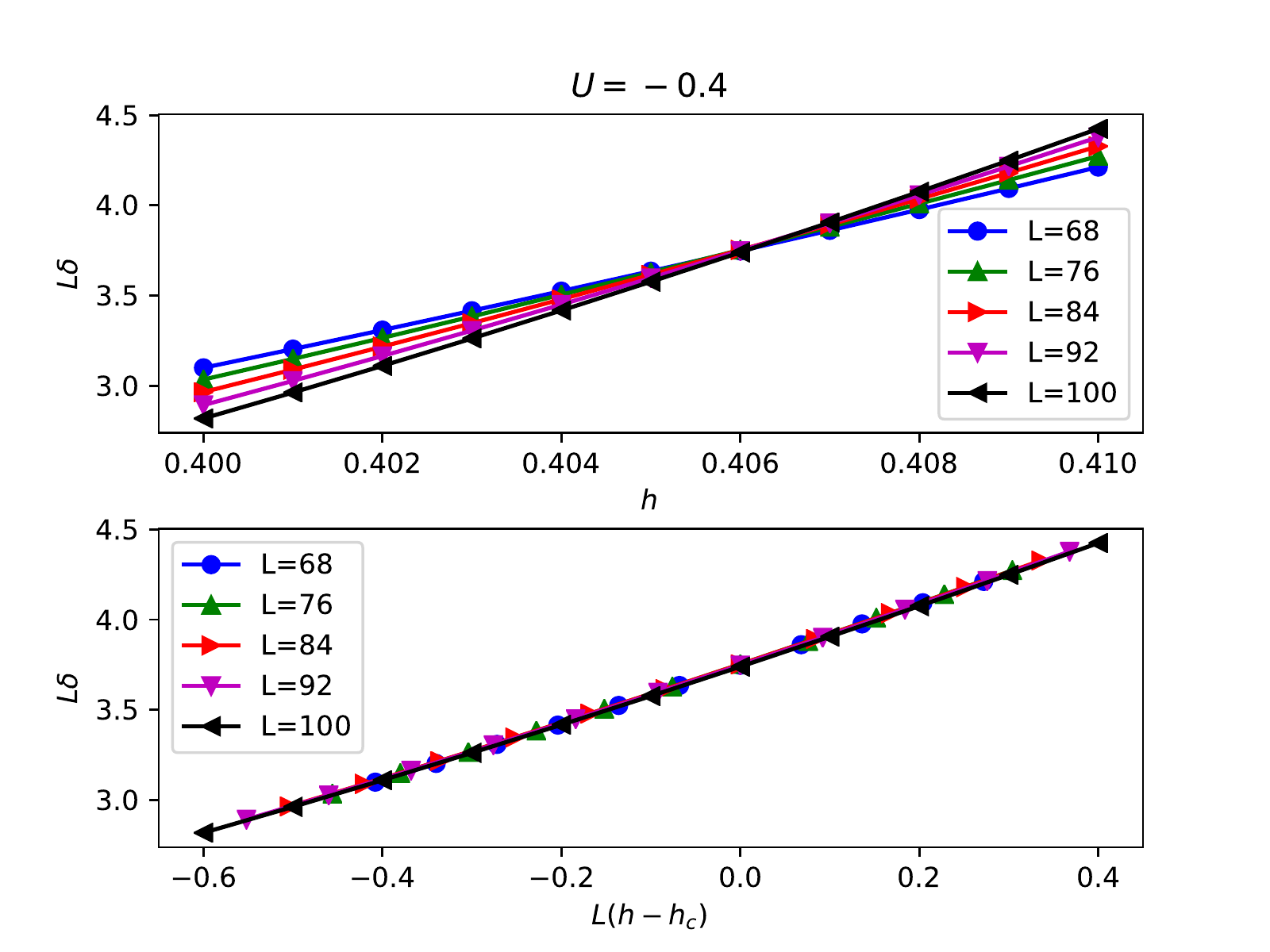}
\caption{(Color online) Finite size scaling for $U=-0.4$. The quantity $L\delta=L(E_1-E_0)$ for different system sizes cross at $h=0.406$.} 
\label{fig-crossing}
\end{figure}

We present our data for $U=-0.4$ as an example. For a given $U$ at $h=h_c(U)$, the quantity $L^z \delta$ does not depend on the system size. In Fig.~\ref{fig-crossing} it is clear that $L \delta$ for different system sizes cross at $h_c=0.406$. One can use this critical value and take $\nu=1$ to check the validity of the ansatz Eq.~\eqref{eq:scaling-form}, see Fig.~\ref{fig-crossing}. Indeed, the data for different system sizes follow the same functional form if one scales the $h$-axis appropriately.    

We used FSS for $-0.8 \leq U \leq 0.25$ and calculated the critical field, $h_c(U)$. The data is presented in Fig.~\ref{fig-bos}. We fitted a power law with the form
 \begin{equation}
 h_c(U)=a(U+1)^b \ , 
 \end{equation}
to the data. Using the data in the range of $-0.8 \leq U \leq -0.5$ the fitting gives $a = 1.02$ and $b =1.81$ . For the full range of data in Fig.~\ref{fig-bos}, however, the best power law fit has $a=0.99$ and $b=1.73$. The blue line in the Fig.~\ref{fig-bos} is the fit to the full set of data points. These results are in a very good agreement with our bosonization result in Eq.~\eqref{eq-result-bosonization}. Our results are also in very good agreement with the perturbative approach of Ref.~\cite{gergs} around the point $h=1$, $U=0$. 

To determine the central charge, $c$, along the curve we used the scaling of the excited states' energy as a function of system size $L$\cite{BYB} and the Calabrese-Cardy (CC) formula for the entanglement entropy (EE) of a subsystem of size $l$, in the case of an open chain at the critical point, $S(l,L)$\cite{cc04,cc09,wilczek94}
\begin{equation}
\label{eq-Cal-Cardy}
S(l)= \frac{c}{6} \log \left( L \sin\left( \frac{ \pi l}{L}\right) \right) +S_0 \ .
\end{equation}
Here $c$ is the central charge and $S_0$ is a constant. Both methods result in the central charge $c = \frac{1}{2}$. Therefore we have the central charge $c=\frac{1}{2}$ all along the transition line except for $h=0$ and $U=-1$, where we have the $XY$ critical point with central charge $c=1$ \cite{BYB}. 

\section{The Repulsive Interaction case}
\label{sec-repulsive}
In this section we study the Hamiltonian Eq.~\eqref{eq-q-int-ising} with repulsive interaction, $U > 0$, which is more involved. We can qualitatively understand the physics in the weak and strong interaction limits. For weak interaction, $U \ll 1$, we still expect to have two phases, i.e. a topological (ordered) phase and a trivial (disordered) phase, as in the case of the TFIM. On the other hand for very strong interaction, the model is in the N{\'e}el phase which corresponds to the charge density wave (CDW) in the fermionic representation. Apart from these two easy limits it is hard to say something about the phase(s) for intermediate interaction strength. For the earlier studies of the model with repulsive interaction using DMRG and bosonization we refer to Refs.~\onlinecite{feo06,feo07,nagy,allen}.

To find the phase diagram of the model we performed DMRG using the ALPS libraries\cite{alps07,alps11,alps14}. We used the fermionic incarnation of the model, i.e. Eq.~\eqref{eq-fermionic-incarnation}, to perform the DMRG. This is useful since the total parity $P$ is a conserved quantity, 
\begin{equation}
P=\prod_{j=1}^L \sigma ^z_j=\prod_{j=1}^L(1-2n_j) \ .
\end{equation}
In above the parity is given in both bosonic and fermionic incarnation. Since $P^2=\mathbf{1}$, its eigenvalues are $P=\pm 1$. Because parity is a good quantum number, the eigenstates with $P=+1$ are called the even sector and the ones with $P=-1$ the odd sector. One can restrict DMRG to be done in one sector\cite{alps14}. 

In a default setup we performed DMRG for a chain of size $L=240$, used the bond dimension $\chi=500$ for the Schmidt decomposition and swept eight times. By the convention in ALPS each sweep means an optimization from the first site to the last one and again from the last site to the first one. We have, however, checked our results to be sure that the DMRG is converged. Specifically, we increased the number of sweeps up to twelve, and considered different system sizes. We already note that in the green region of the phase diagram (Fig.~\ref{fig-pd}), the system size modulo four plays an important role.
We discuss this issue extensively below. 

\begin{figure}[h]
\includegraphics[width=\columnwidth]{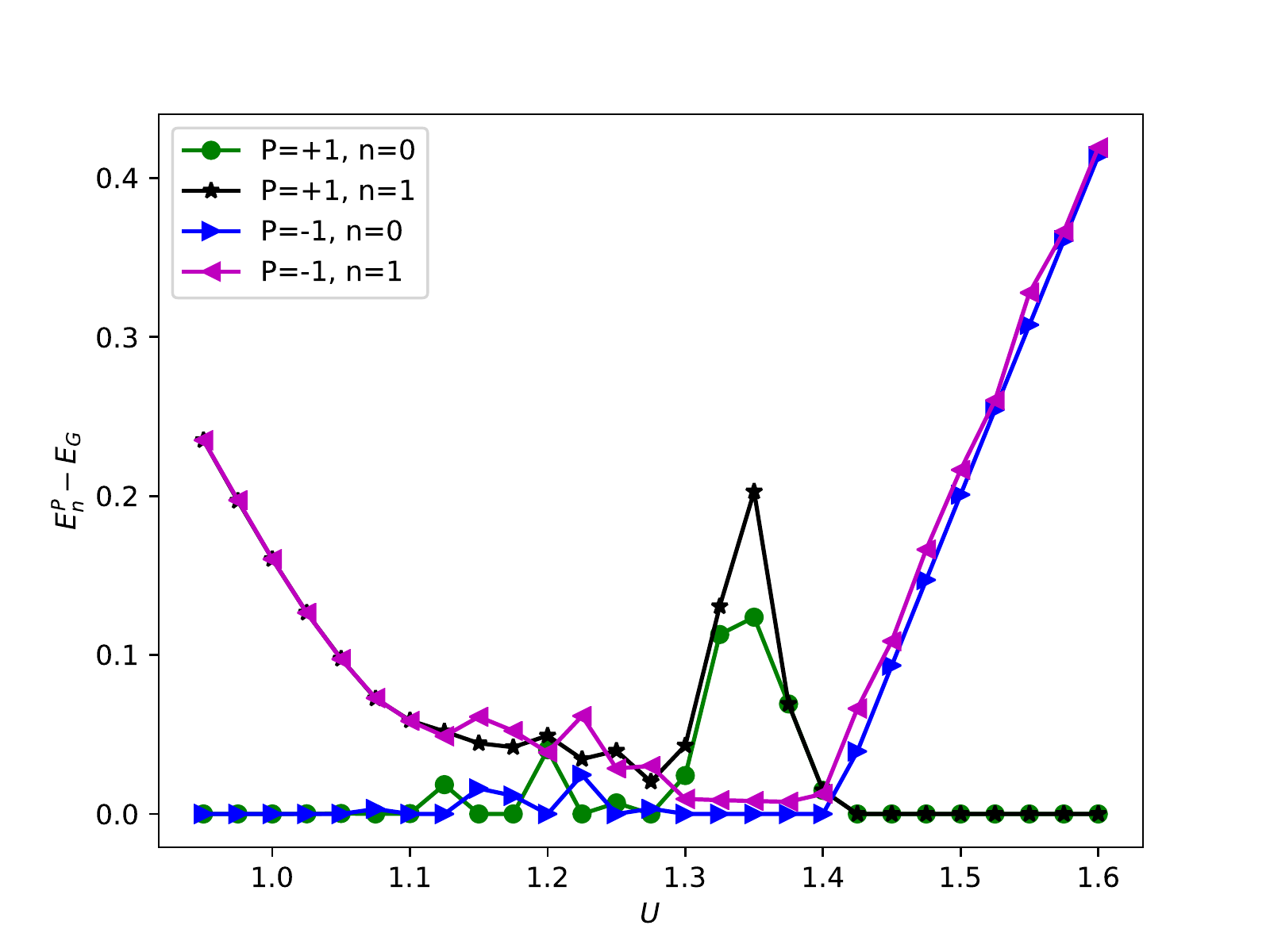}
\caption{(Color online) Energy with respect to {\it the ground state energy}, $E^{P}_{n} - E_G$, of the ground state ($n=0$) and the first excited state ($n=1$) in both parity sectors are plotted as function of $U$ for $h=0.7$ and $L=240$. Note that $E_G$ is either $E^{+1}_{0}$ or $E^{-1}_{0}$. The range of $U$ corresponds to the dashed line in Fig.~\ref{fig-pd}.} 
\label{fig-four_states}
\end{figure}

To begin our tour of the repulsive regime, we present our results for the lowest two eigenstates in each parity sector, the ground state ($n=0$) and the first excited state ($n=1$). We label the states with $P$ and $n$ and their energy will be denoted by $E^P_n$. The result for $h=0.7$ along the dashed line in Fig.~\ref{fig-pd} is presented in Fig.~\ref{fig-four_states}. One can see features of different phases in this plot. For $U \lesssim 1.05$ the model is in a gapped topological phase in which the ground state is doubly degenerate, with opposite parities.

For $1.05 \lesssim U \lesssim 1.29$ we have a gapless incommensurate phase in which observables oscillate and the parity of the ground state depends highly on details such as the system size and the couplings. Increasing $U$ further, for $1.29 \lesssim U \lesssim 1.41$ we have a new phase which we believed has been missed previously. The intriguing feature of this region is that the nature of the ground state differs for even and odd system sizes. Finally for $U \gtrsim 1.41$ the system has a CDW ground state. We will later discuss the degeneracy and the parity of the ground state in this phase for different system sizes.

\subsection{The topological phase} 

\begin{figure}[htp]
\subfloat[Entanglement entropy as a function of subsystem size.]{\includegraphics[clip,width=\columnwidth]{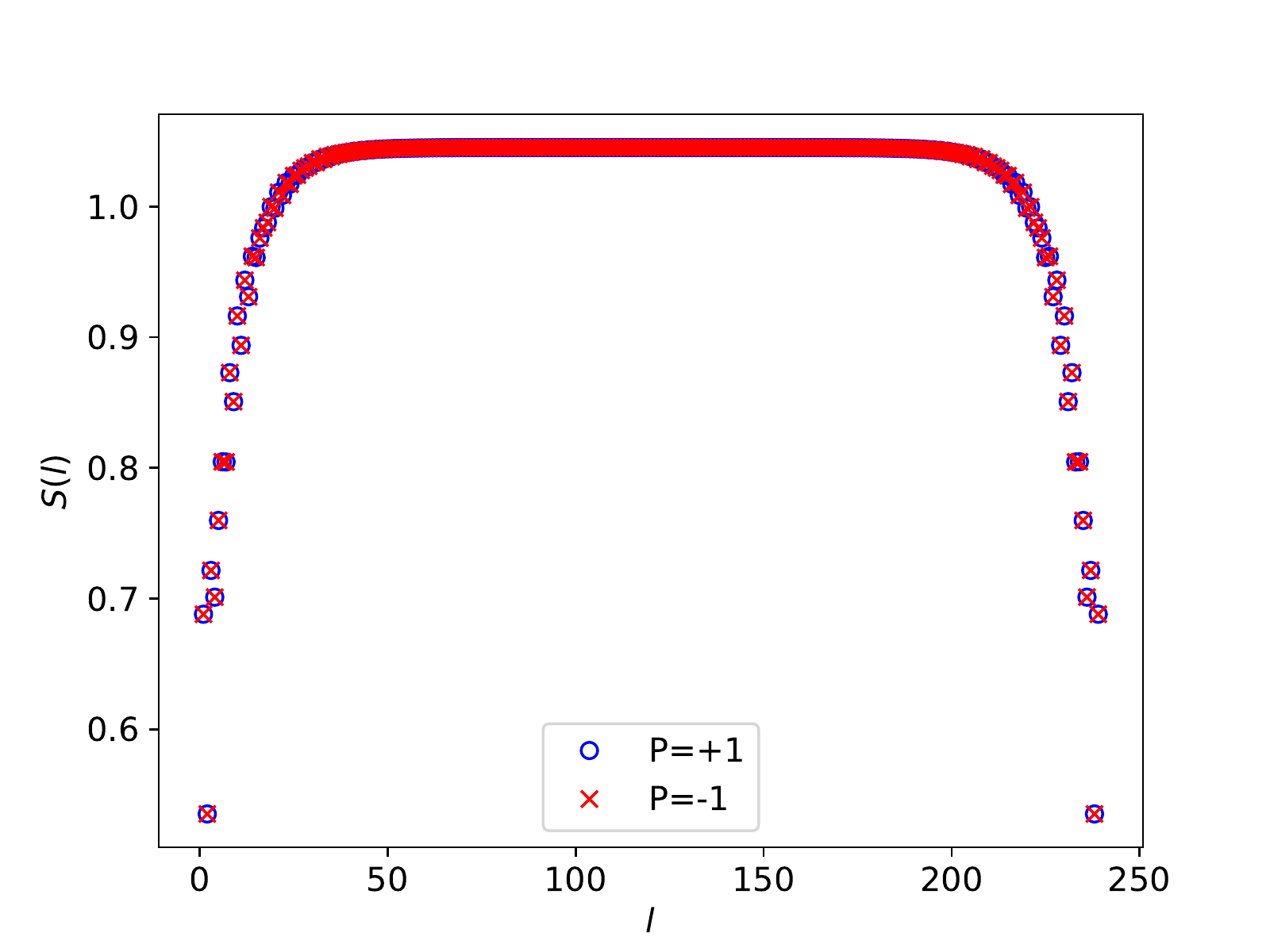}}

\subfloat[Occupation number in the ground state.]{ \includegraphics[clip,width=\columnwidth]{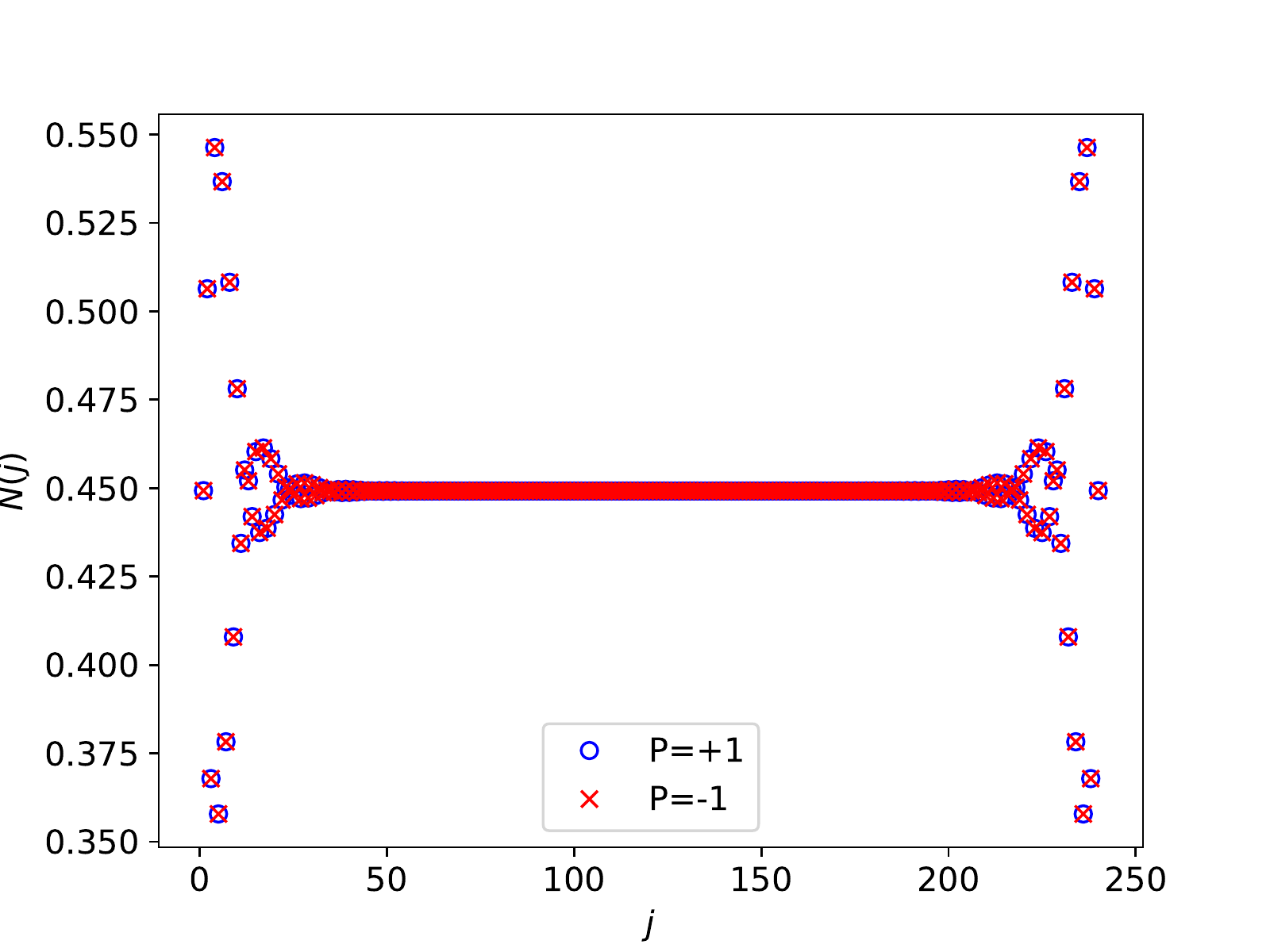}}
\caption{Observables for $h=0.7$ and $U=0.95$, a point in the topological phase, are plotted for the two ground states. The system size is $L=240$.}
\label{fig-topo_sample}
\end{figure}

In Fig.~\ref{fig-four_states} we start in the topological phase (the white region in Fig.\ref{fig-pd}), where the model is gapped and the ground state is doubly degenerate. The two ground states have different parity. In Fig.~\ref{fig-topo_sample} we present the EE, $S(l)$, as a function of subsystem size, $l$, and the (expectation value of the fermion) occupation number at each site, $N(j)$, for the ground state in each parity sector. The difference of the energy of the two states is of the order of $10^{-7}$ in units of hopping strength, note that we set $t=1$ in Eq.~\eqref{eq-fermionic-incarnation}. As is evident, $S(l)$ saturates to a high value in the bulk for both the ground states, which is a signature of the topological phase. The occupation number is constant in the bulk for both parity sectors. 

\subsection{The incommensurate phase}
\label{sec-on_ic_phase}

By increasing the interaction strength we enter the incommensurate (IC) phase, the orange region in Fig.~\ref{fig-pd}. In part of the literature, especially papers considering the classical model or the dual model, this phase is called the floating phase. In this phase the ground state is unique but its parity does strongly depend on the parameters and the system size. 

\begin{figure}[htp]
\subfloat[Entanglement entropy as a function of subsystem size.]{\includegraphics[clip,width=\columnwidth]{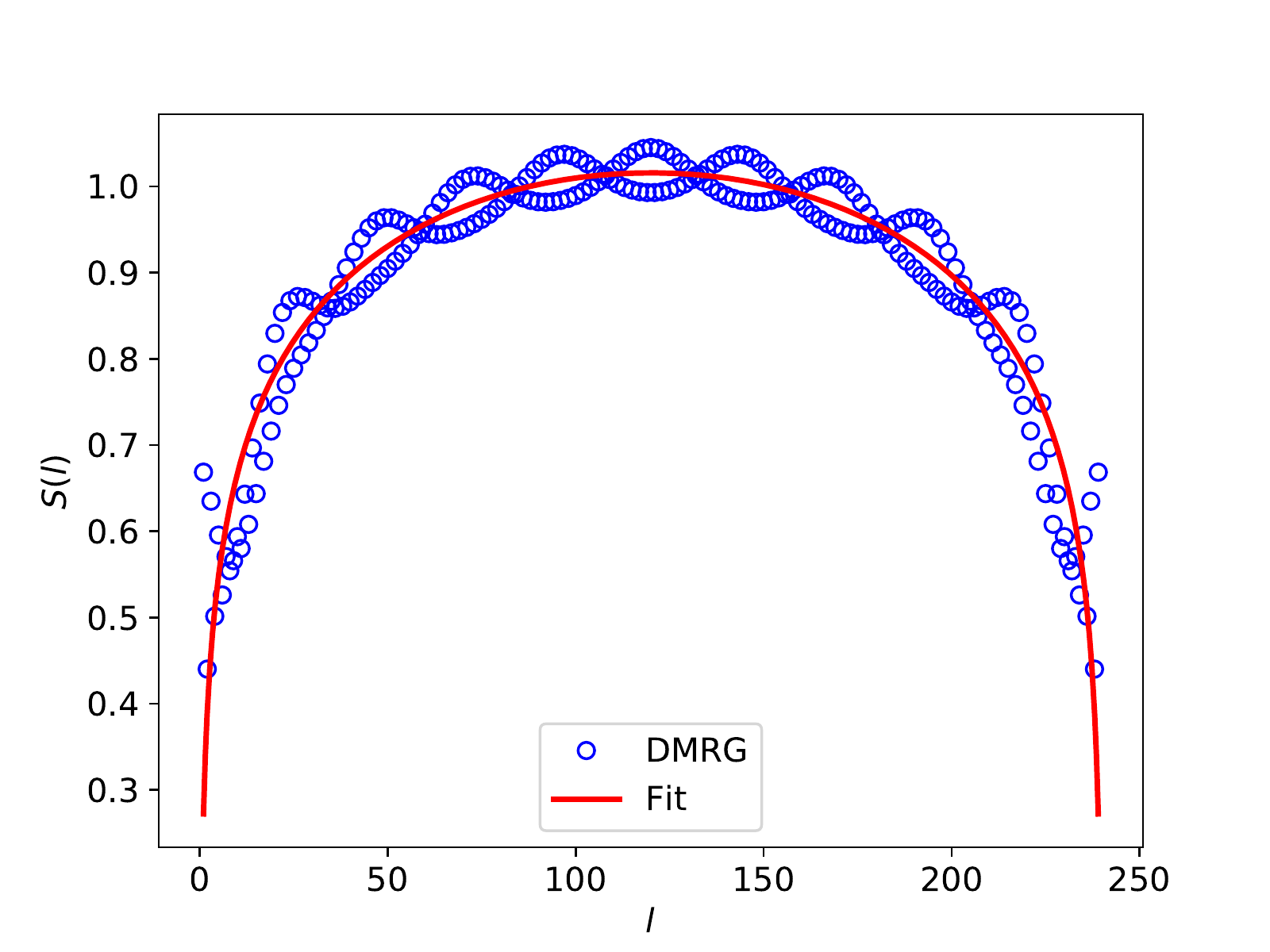}}

\subfloat[Occupation number in the ground state.]{ \includegraphics[clip,width=\columnwidth]{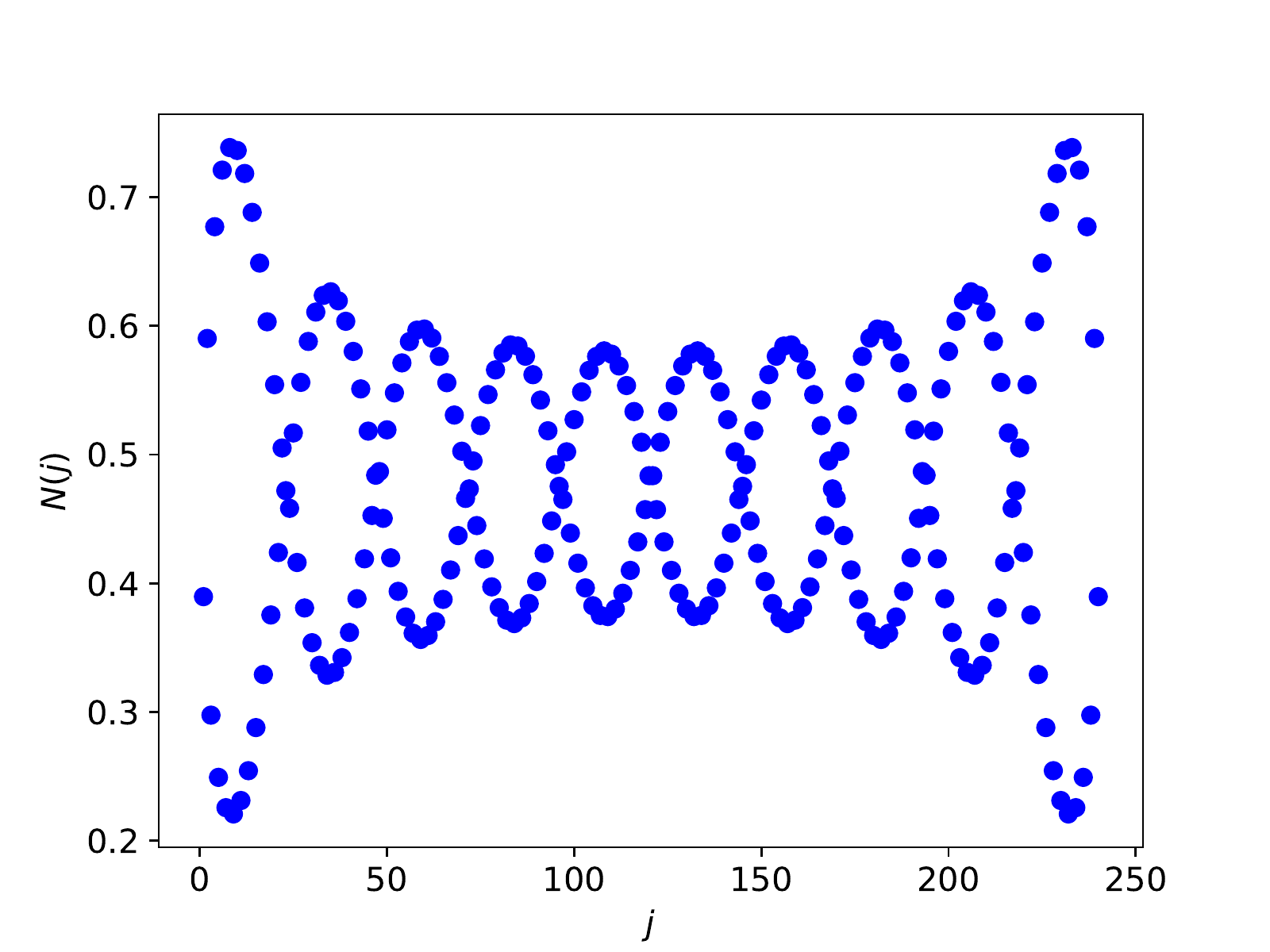}}
\caption{Ground state observables for $h=0.7$ and $U=1.2$, a point in the incommensurate phase, are plotted for $L=240$. The ground belongs to the odd sector.}
\label{fig-IC_sample}
\end{figure}

In Fig.~\ref{fig-IC_sample} we present the EE and occupation number $N(j)$ for a chain of size $L=240$ at a point in the IC phase, $h=0.7$ and $U=1.2$. For this set of parameters the ground state happens to belong to the odd sector.

One signature of this phase is the presence of oscillations in the EE and occupation number. Another feature of this phase is that one can use the CC formula, Eq.~\eqref{eq-Cal-Cardy}, for the EE to fit the data. Previously this has been done to distinguish the IC phase from the topological and the trivial phases of a $\mathbb{Z}_3$ parafermionic chain\cite{hughes15}.  

For fitting we usually drop the first and the last $20$ sites because there are clear edge and finite size effects in the both EE and occupation number, see Fig.~\ref{fig-IC_sample}. Using the CC formula, Eq.~\eqref{eq-Cal-Cardy}, to fit the data, we get $c = 1.03$. The fitted curve is also plotted in Fig.~\ref{fig-IC_sample} as a solid line. Hence the essential features of the EE in the IC phase can be captured by the CC formula with $c=1$. One would, of course, get a better fitting by considering the fluctuations on top of the CC formula, but we did not attempt this. We checked that the gap within each parity sector scales as $1/L$, consistent with the fact that the main features of the EE are well described by the CC formula.

Pinpointing the transition from the topological phase to the IC phase is rather a hard task and this transition was conjectured to be of Kosterlitz-Thouless type\cite{selke,feo07}. We looked at the energy and its derivatives and we didn't find any signature of a first or second order transition. We set the transition point by the energy difference of the order of $10^{-3}$ between the ground state in the two parity sectors and by requiring $c=1$ for fitting the CC formula to the DMRG data. Note that for a point in the topological phase close to the transition the EE can also be fitted well to the CC formula with central charge $c<1$.  One can distinguish the topological phase from the IC phase by looking at larger system sizes. In the topological phase, for large enough system size we do see the saturation of EE, however, in the IC phase one would obtain $c=1$ also for larger system sizes.

Another qualitative check for distinguishing the IC phase is the presence of oscillations in the occupation number. This can be verified by looking at the bulk of the chain, compare the (b) panels of Figs.~\ref{fig-topo_sample} and \ref{fig-IC_sample}.

From Fig.~\ref{fig-four_states} it is evident that the parity of the ground state in the IC phase changes. From this figure one can see that there are several crossings between the ground states of the two parity sectors, i.e. $E^{+1}_0$ and $E^{-1}_0$. One should keep in mind that both the precise point where the these crossings happen, as well the parity of the ground state in this region depends on the system size. Nevertheless there are two important crossings, namely the last crossing at $U\simeq 1.41$ and the next-to-last crossing at $U \simeq 1.28$. We will later discuss the region between these two crossings in detail in Sec.~\ref{sec:new-phase}. Note that for $h=0.7$ the IC phase ends at the next-to-last crossing at $U \simeq 1.28$ in Fig.~\ref{fig-four_states}.  

\begin{figure}[h]
\includegraphics[width=\columnwidth]{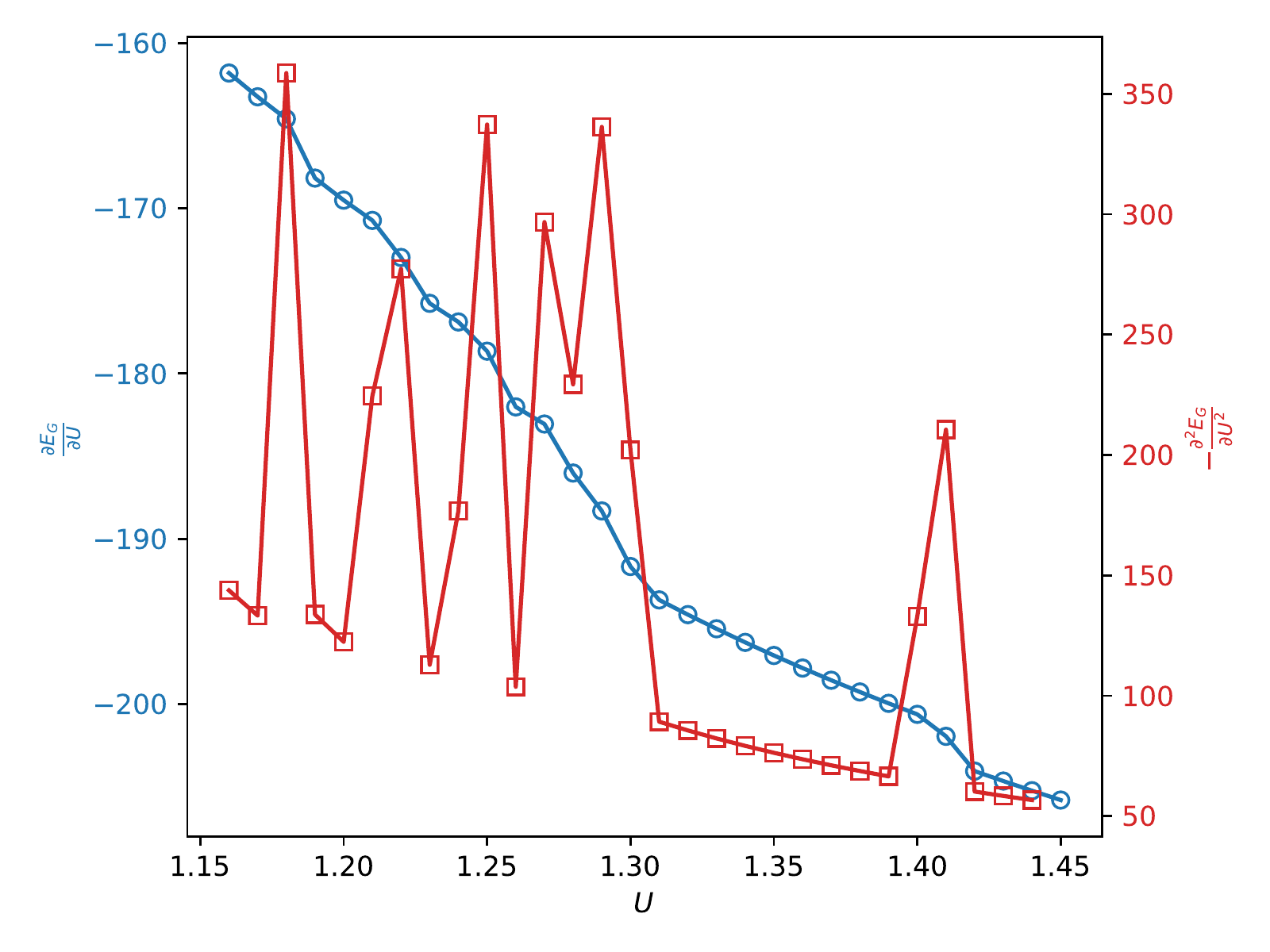}
\caption{(Color online) The derivatives of the ground state energy, $\frac{\partial E_G}{\partial U}$ (the blue circles) and $-\frac{\partial^2 E_G}{\partial U^2}$ (the red squares) are plotted against $U$ for $h=0.7$.} 
\label{fig-ic-cdw}
\end{figure}

To verify that level crossings occur we looked at the ground state energy and its derivatives with the steps of $10^{-2}$ along the $U-$axis. The first and the second derivatives of the ground state energy with respect to $U$ are shown in Fig.~\ref{fig-ic-cdw}. There are discontinuities in $\frac{\partial E_G}{\partial U}$ which give rise to the peaks in $-\frac{\partial^2 E_G}{\partial U^2}$. We have looked at the similar quantities for different system sizes and the discontinuities are always present in the first order derivative. These are clear signatures of level crossings. We note that both the number of crossings, as well as their location, depends rather strongly on the system size. 

To address the question about the presence of the floating phase in the ANNNI model (Eq.~\eqref{eq-qannni}) in the highly frustrated limit, i.e. $\kappa \gg 1$, we can use our knowledge of the Kitaev-Hubbard chain. Because the Kitaev-Hubbard chain is dual to the ANNNI model, the presence of the IC phase in the fermionic incarnation corresponds to the presence of the floating phase in the ANNNI model. 

\begin{figure}[htp]
\subfloat[Energy with respect to {\it the ground state energy}, $E^{P}_{n} - E_G$, of the ground state ($n=0$) and the first excited state ($n=1$) in both parity sectors are plotted as function of $U$ for $h=0.05$.]{\includegraphics[clip,width=\columnwidth]{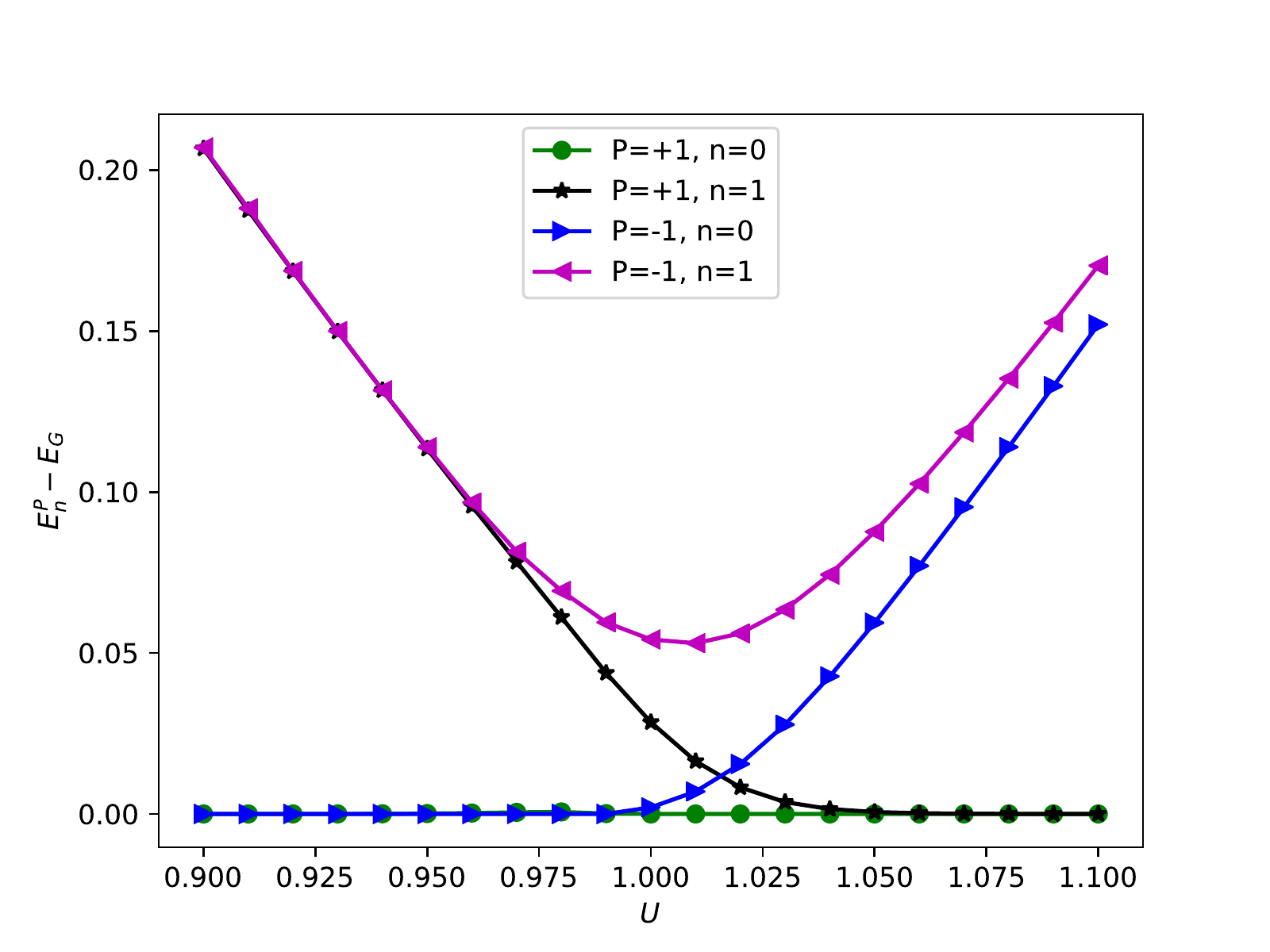}}

\subfloat[The derivatives of the ground state energy, $\frac{\partial E_G}{\partial U}$ (the blue circles) and $-\frac{\partial^2 E_G}{\partial U^2}$ (the red squares) for $h=0.1$ are plotted.]{ \includegraphics[clip,width=\columnwidth]{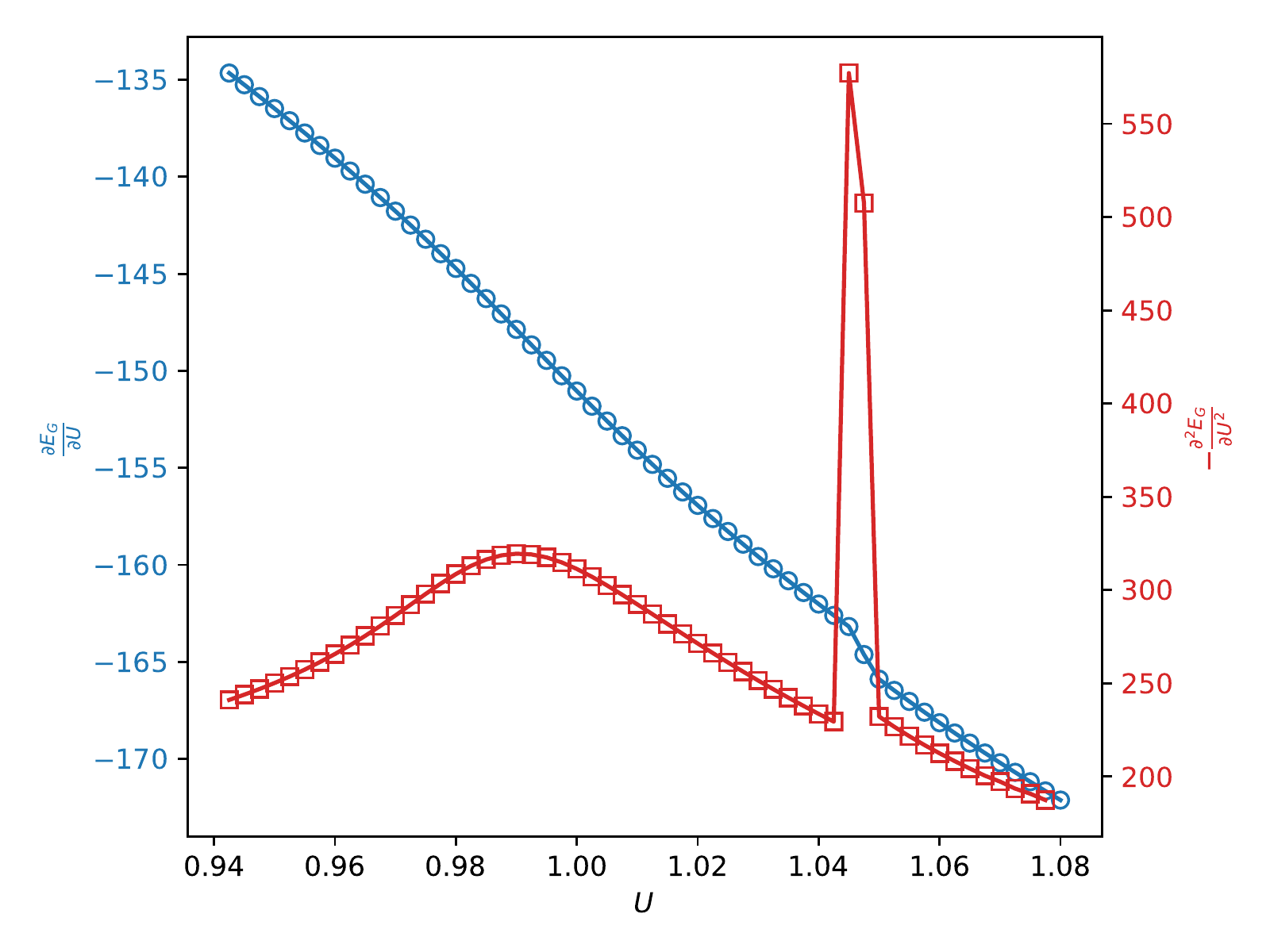}}
\caption{Energy levels and the derivatives of the ground state energy for small $h$ are plotted. The system size is $L=240$. }
\label{fig-samll_h}
\end{figure}

From Eq.~\eqref{eq-parameters_dualirty} we can see that for $U \simeq 1$ and small $h$, we will be in the regime of $\frac{B}{J_1}, \kappa \gg 1$, which is exactly the regime where the presence of the floating phase in the ANNNI model was under debate. In Fig.~\ref{fig-samll_h}~(a) we plot the energy of the ground state $(n=0)$ and the first excited state $(n=1)$ in both parity sectors for a chain of length $L=240$, at $h=0.05$. This corresponds to $\kappa \simeq 20$. The situation is similar to $h = 0.7$, as plotted in Fig.~\ref{fig-four_states}. For low enough $U$ the ground state is doubly degenerate and the states have different parities, i.e. a topological phase. For large enough $U$, the ground state is also degenerate and the states have the same parity, i.e. $P=+1$ in this case. 

In Fig.~\ref{fig-samll_h} (b), one observes a discontinuity in $\frac{\partial E_G}{\partial U}$ and the corresponding peak in $-\frac{\partial^2 E_G}{\partial U^2}$ for $h=0.1$ (which corresponds to $\kappa=10$) at $U = 1.05$, as well as a broad feature at $U=0.99$.
In addition, it is also possible to fit the EE with the CC formula, resulting in $c=1$ for a small region in $U$, even for $h=0.05$. Finally, one observes an onset to the oscillations in the occupation number, that is characteristic for the IC phase. Given these features, we conclude that the IC phase in the Kitaev-Hubbard chain continues down to the $XY$ critical point. Therefore, we also conclude that the floating phase in the ANNNI model is present for arbitrarily large frustration.

\subsection{The excited-state CDW phase}
\label{sec:new-phase}

In our numerical studies we found that the ground state for $h = 0.7$ and $ U \gtrsim 1.41$ has CDW order and a low EE, which is constant in the bulk, that is, independent of the subsystem size. By looking at the energy levels in Fig.~\ref{fig-four_states} we see that there is a region, $1.28 \lesssim U \lesssim 1.41$, just before the CDW phase in which the the ground state is in the odd sector for our default system size, $L=240$. We believe that this part of the phase diagram, coloured light green in Fig.~\ref{fig-pd}, is a phase which had been missed before and has been considered to be part of the CDW. We will discuss the properties of this phase thoroughly and show how the behaviour of the model in this part of the phase diagram depends on the system size. Moreover we show that the transition point from the new phase to the CDW phase can be actually controlled by tuning the chemical potential on the first and the last sites.

\begin{figure}[htp]
\subfloat[Entanglement entropy as a function of subsystem size.]{\includegraphics[clip,width=\columnwidth]{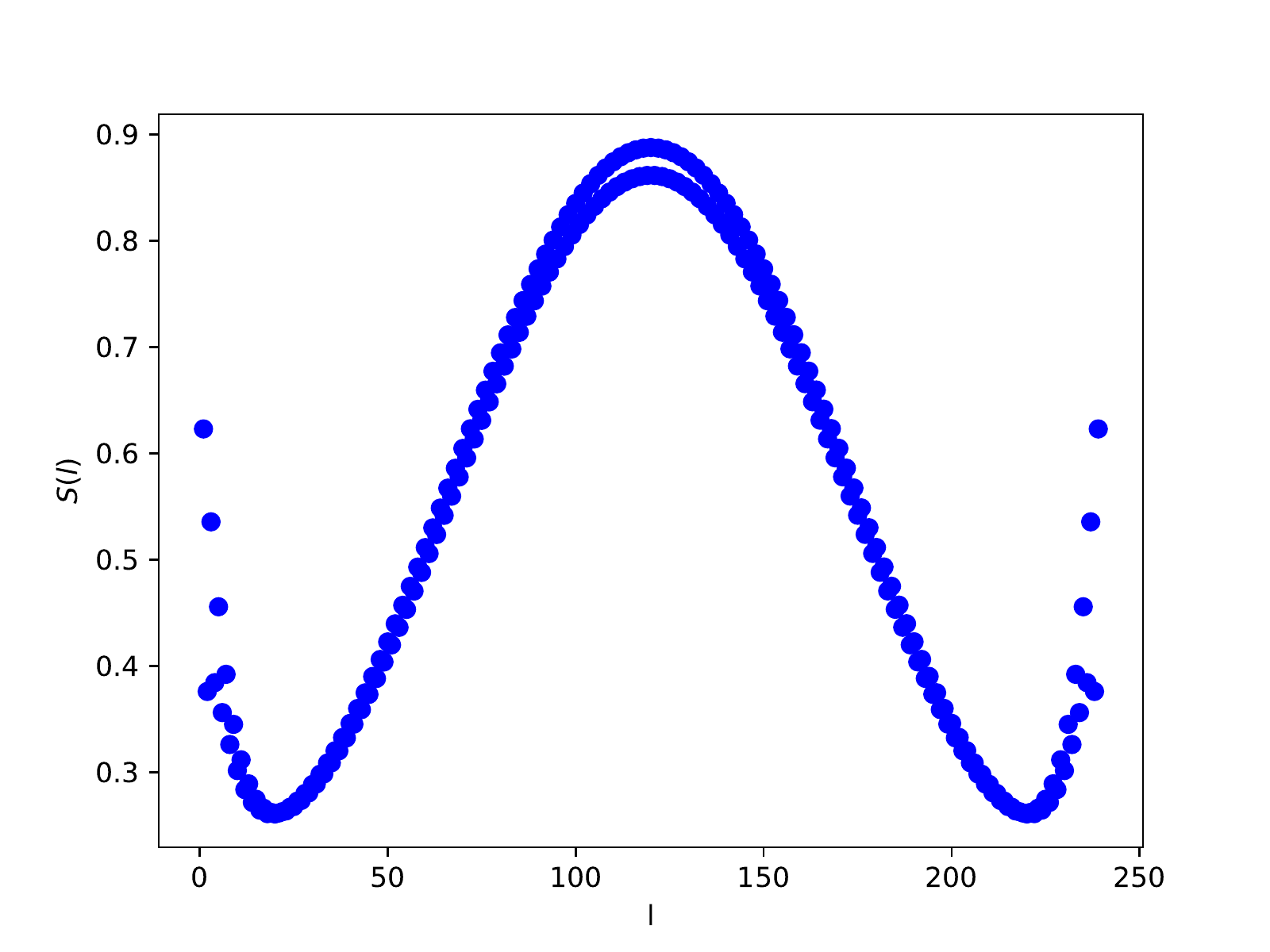}}

\subfloat[Occupation number.]{ \includegraphics[clip,width=\columnwidth]{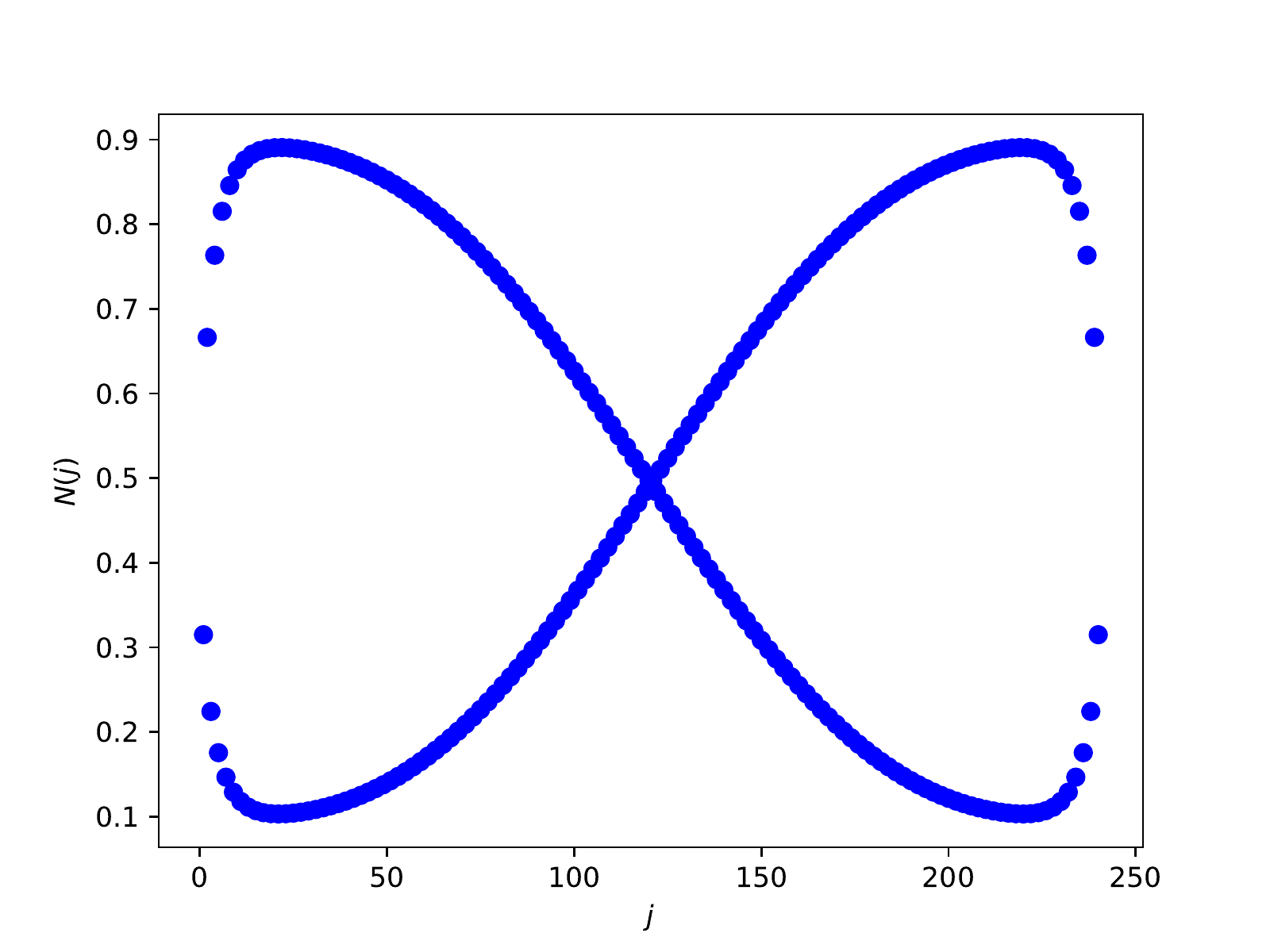}}
\caption{Observables for the ground state at $h=0.7$ and $U=1.325$, a point in the esCDW phase are plotted. The system size is $L=240$.}
\label{fig-new_phase_sample}
\end{figure}

\begin{figure}[h]
\includegraphics[width=\columnwidth]{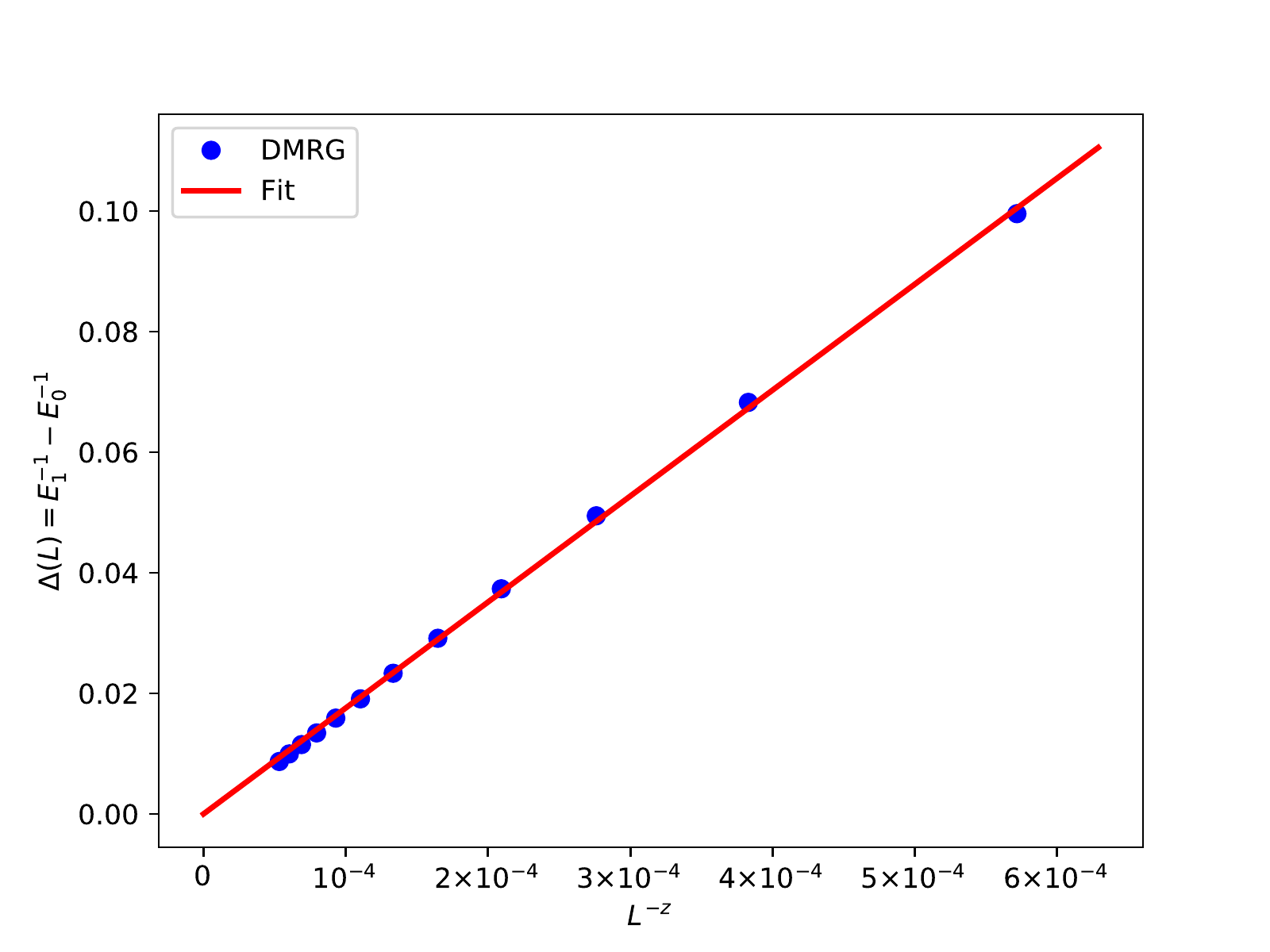}
\caption{Scaling of the gap, $E^{-1}_{1} - E^{-1}_{0}$, for systems of the size $L=4n$ in the range of $L=64-240$ at $h=0.7$ and $U=1.325$.} 
\label{fig-new_phase_gap_scaling}
\end{figure}

\begin{figure}[h]
\includegraphics[width=\columnwidth]{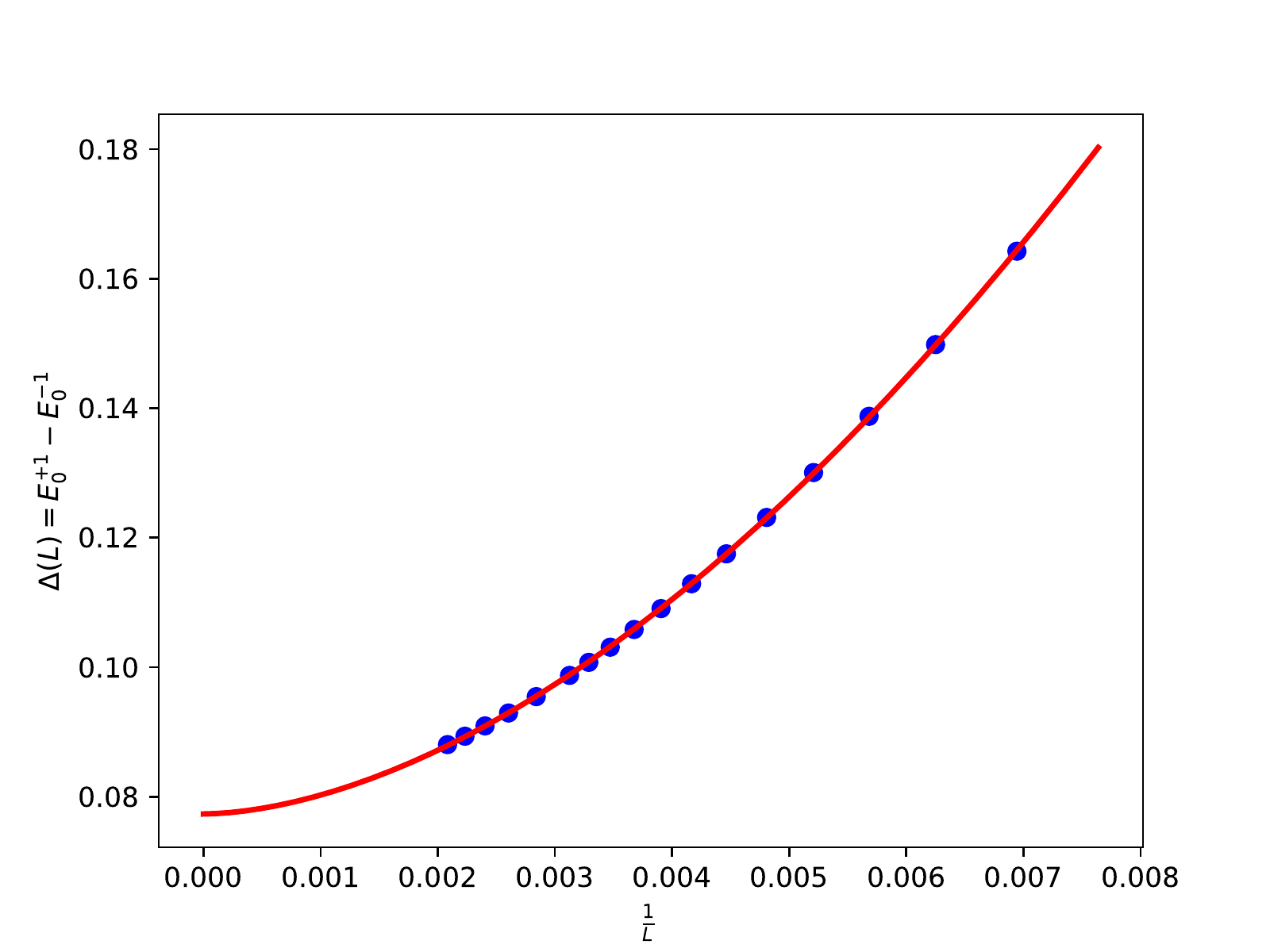}
\caption{Scaling of the gap, $E^{+1}_{0} - E^{-1}_{0}$, for systems of the size $L=4n$ in the range of $L=144-480$ at $h=0.7$ and $U=1.325$. There is a finite gap,$\Delta \simeq 0.077$, between the ground states of the two sectors.} 
\label{fig-gap_even_ecdw}
\end{figure}

We present the EE and occupation number for a generic point ($h = 0.7$ and $U = 1.325$) in the new phase in Fig.~\ref{fig-new_phase_sample}.
It is clear the EE of the ground state grows as a function of subsystem size. The particle number has a long wavelength oscillation of twice the system size, accompanied by a $\pi$ phase shift for neighbouring sites. These observations show that the ground state does not have a conventional CDW ordering for our system size. In fact, the properties of the ground state resemble the properties of low-lying excited states in the CDW phase. We therefore refer to this phase as the `excited-state charge density wave' (esCDW) phase.
In addition, we found that the model is gapless by looking at the first ($n=1$) and the second ($n=2$) excited states in the odd sector. The energy difference between them, $\Delta E^{-1}_n = E^{-1}_{n} - E^{-1}_{0}$, approaches zero as $L^{-1.80}$ for $n=1,2$. The data for $n=1$ and the fit are shown in Fig.~\ref{fig-new_phase_gap_scaling}.
This shows this phase is a gapless phase with the dynamical critical exponent $z\simeq 1.80$ and hence can not be described by a conformal field theory. We also found that there is a finite gap to the lowest energy state in the even sector. In Fig.\ref{fig-gap_even_ecdw} we plot $E^{+1}_{0} - E^{-1}_{0}$ as a function $\frac{1}{L}$ which shows the existence of the gap.

An intriguing feature of this region of the phase diagram is that its features depend crucially on the system size. The ground state belongs to the odd sector for system sizes that are a multiple of four, $L=4n$. For $L=4n +2$ however, the ground state shows the same behaviour as described above, but it belongs to the even sector.

\begin{figure}[h]
\includegraphics[width=\columnwidth]{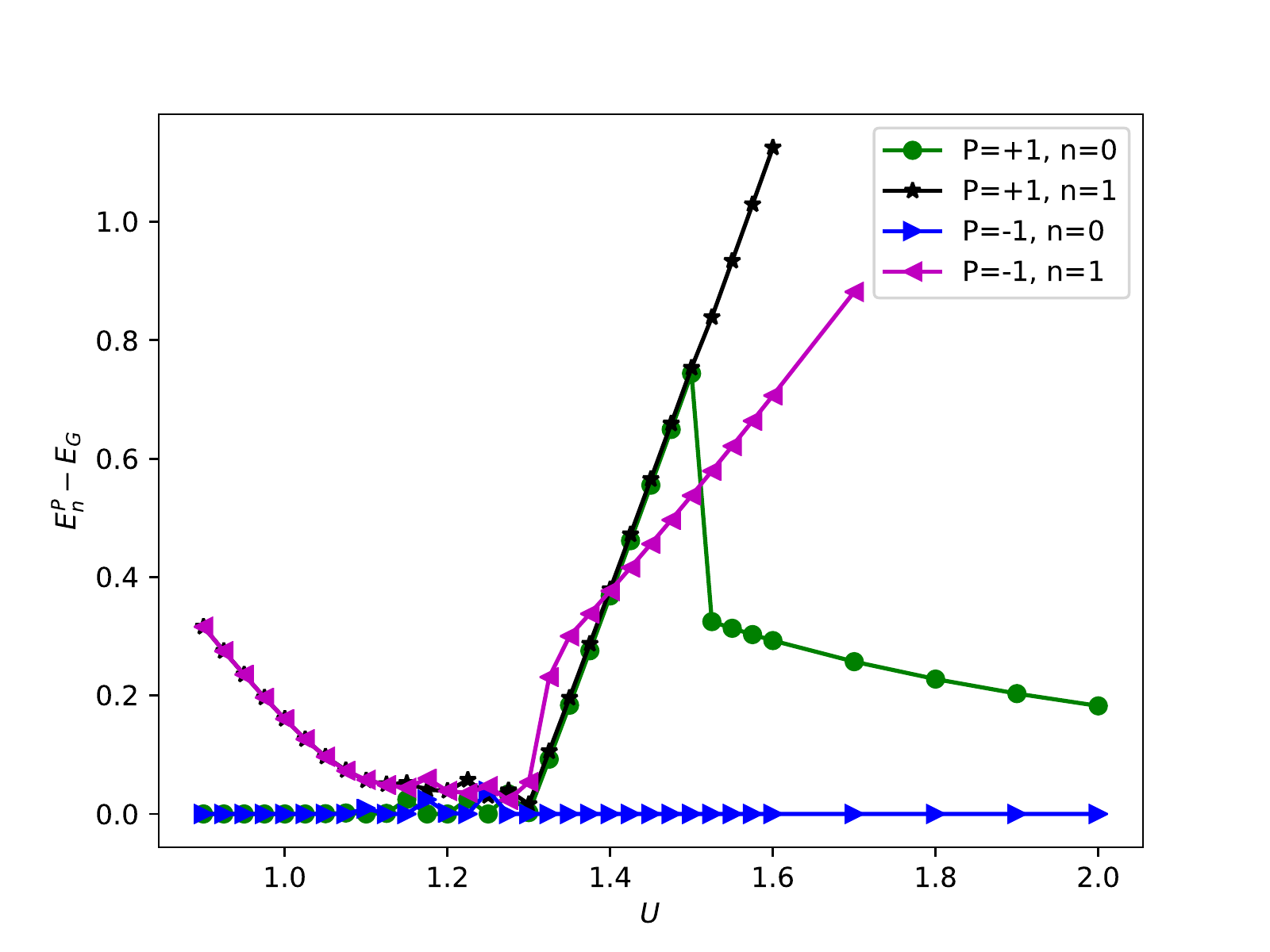}
\caption{Energy with respect to {\it the ground state energy}, $E^{P}_{n} - E_G$, of the ground state ($n=0$) and the first excited state ($n=1$) in both parity sectors are plotted as function of $U$ for $h=0.7$ and $L=243$. Note that $E_G$ is either $E^{+1}_{0}$ or $E^{-1}_{0}$. The range of $U$ corresponds to the dashed line in the Fig.~\ref{fig-pd}.} 
\label{fig-L_odd_energy_cut}
\end{figure}

For an odd number of sites, $L= 4n + 1$ (or $L= 4n + 3$), however, the model is gapped and has a unique ground state with the CDW ordering in the even (odd) sector. In Fig.~\ref{fig-L_odd_energy_cut} we present the two lowest states in each parity sector for $L=243$. One should compare this plot with Fig.~\ref{fig-four_states}; one observes that after the IC phase there is only one phase. We checked that the ground state has a low EE and the occupation number showed CDW ordering. However, there is one important difference with the CDW phase for even system sizes.
Namely, in Fig.~\ref{fig-four_states} we see that as soon as one enters the CDW phase the ground state is doubly degenerate and both of them have the same parity, i.e. $P=+1$. For odd number of sites, however, the ground state is unique, with a finite gap to the lowest excited state, which has opposite parity. This excited state also has a low EE and CDW ordering. For large $U$, the gap between these states goes to zero, which we verified by considering values for $U$ up to $U \sim 10$. We conclude that for odd system sizes, upon increasing $U$, the IC phase gives way to a CDW phase (with a unique ground state) immediately.

We should comment on one more feature in Fig.~\ref{fig-L_odd_energy_cut}. The energy of the ground state in the even sector (the green dots), makes a rather large jump between $U=1.5$ and $U=1.525$. We thoroughly checked that this jump is not an artifact of the DMRG algorithm. First of all, the states are clearly converged, because the observables all have smooth profiles (as a function of the site number), and the energy does not change if the number of sweeps is increased. In addition, this feature is present for all system sizes and parameters in this region, it does not go away upon slightly varying these parameters.  

\begin{figure}[h]
\includegraphics[width=\columnwidth]{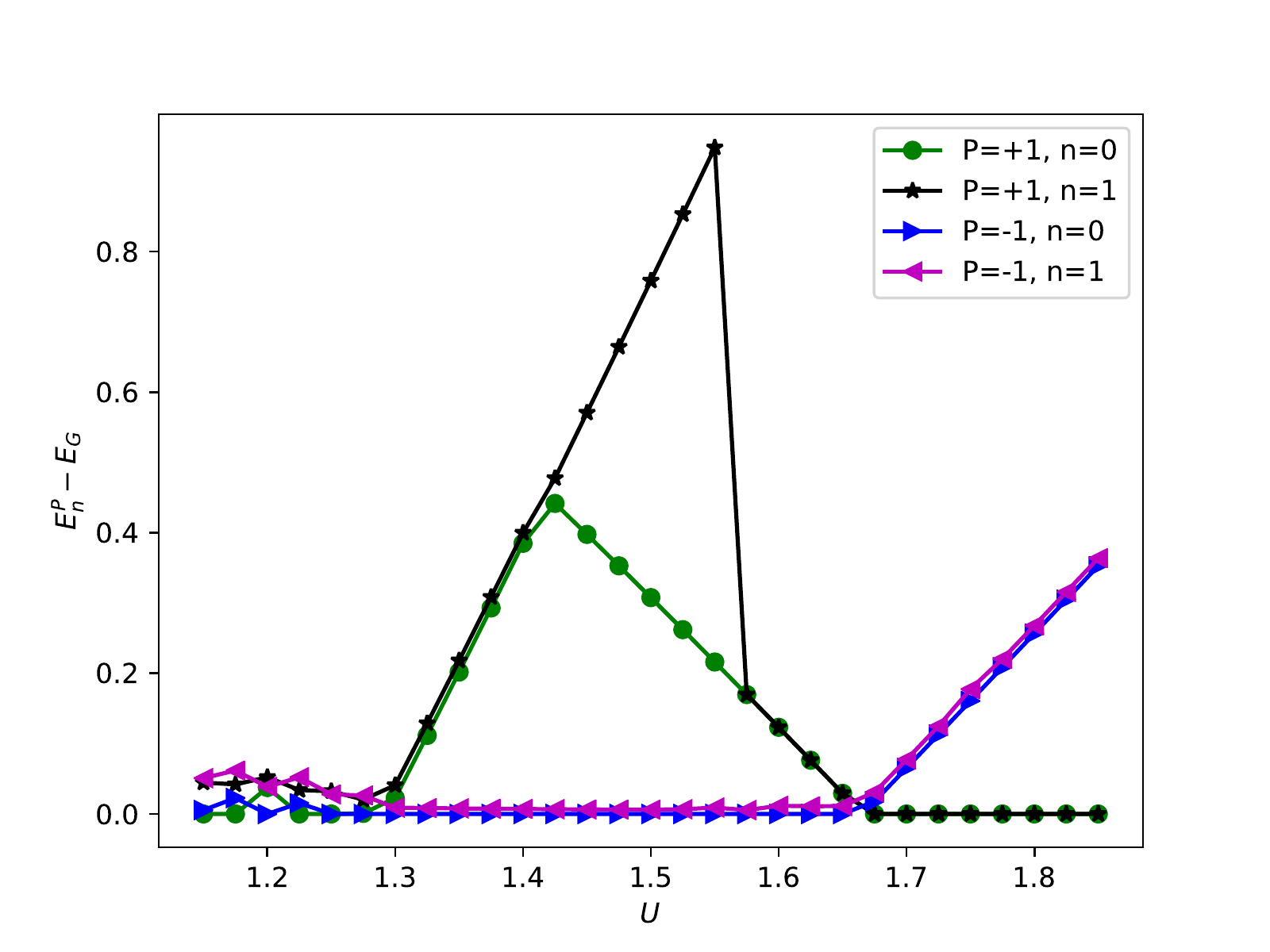}
\caption{Energy with respect to {\it the ground state energy}, $E^{P}_{n} - E_G$, of the ground state ($n=0$) and the first excited state ($n=1$) in both parity sectors are plotted as function of $U$ for $h=0.7$ and $L = 240$ with the implementation of the magnetic field in Eq.~\eqref{eq-chemical-sites}.} 
\label{fig-h_on_sites}
\end{figure}

In the region where the esCDW phase is observed, the behaviour of the model does not only depend on the (parity of the) system size, but also on how one implements the magnetic field term. 
As we mentioned before we have implemented the magnetic field term on bonds, see Eq.~\eqref{eq-chemical-bonds}. We also considered the model with
the same magnetic field on all sites, namely by using the following term in the fermionic incarnation in DMRG, 
\begin{equation}
\label{eq-chemical-sites}
\sum _{j=1}^{L} - h  \left( 1-2 c^{\dagger}_j c_j \right) \ .
\end{equation}
As expected, we did not seen any difference between the two ways of implementing the magnetic field in the topological and the IC phases.
The IC phase for $h=0.7$ still ends around $U\simeq 1.28$, however, the esCDW phase extends up to a larger value of $U$, namely $U \simeq 1.68$, as can
be observed by comparing Figs.~\ref{fig-four_states} and \ref{fig-h_on_sites}, in which we plot the two
lowest energies for both parity sectors as a function of $U$.

Therefore the esCDW phase is sensitive to both the system size as well as the chemical potential on the first and the last sites. To be sure that these results are not due to convergence issues and unluckily chosen system sizes, we checked the results for various smaller system sizes as well, leading to identical results.

We should make two important remarks, before closing our discussion about the esCDW phase. First of all we should mention that if one uses infinite DMRG to study this phase, one would conclude that it belongs to the CDW phase\cite{nagy}. This is due to the fact that the ground state energy per unit site (or per bond) for the odd number of sites is lower than the one for even number of sites by $10^{-3}$ in the units of hopping energy.

Another issue concerns the presence of the esCDW phase for small $h$. We clearly see the presence of the esCDW phase for $h=0.1$. For smaller $h$, however, it is hard to draw firm conclusions.  As we discussed in Sec.~\ref{sec-on_ic_phase} we see the signatures for the crossing between the energy levels and hence discontinuities in the first order derivative of the energy with respect to $U$. However to conclude that both the IC and the new phase are present for small $h$ we need to see two crossings, one which corresponds to the transition from the IC phase to the new phase and another one for the transition from the new phase to the CDW phase. With current accuracy and system sizes we only see one crossing. Because the ground state EE is consistent with the CC formula for $c=1$ we conclude that at least the IC phase is present. However, the the width of these two phases shrinks as we get closer to the $XY$ critical point and distinguishing them becomes harder and needs larger system sizes and more accurate treatments. We expect, however, that the new phase also extends to $h=0$, i.e. down to the $XY$ critical point, as indicated in the phase diagram, Fig.~\ref{fig-pd}.

\subsection{The charge-density-wave phase}

\begin{figure}[htp]
\subfloat[Entanglement entropy as a function of subsystem size.]{\includegraphics[clip,width=\columnwidth]{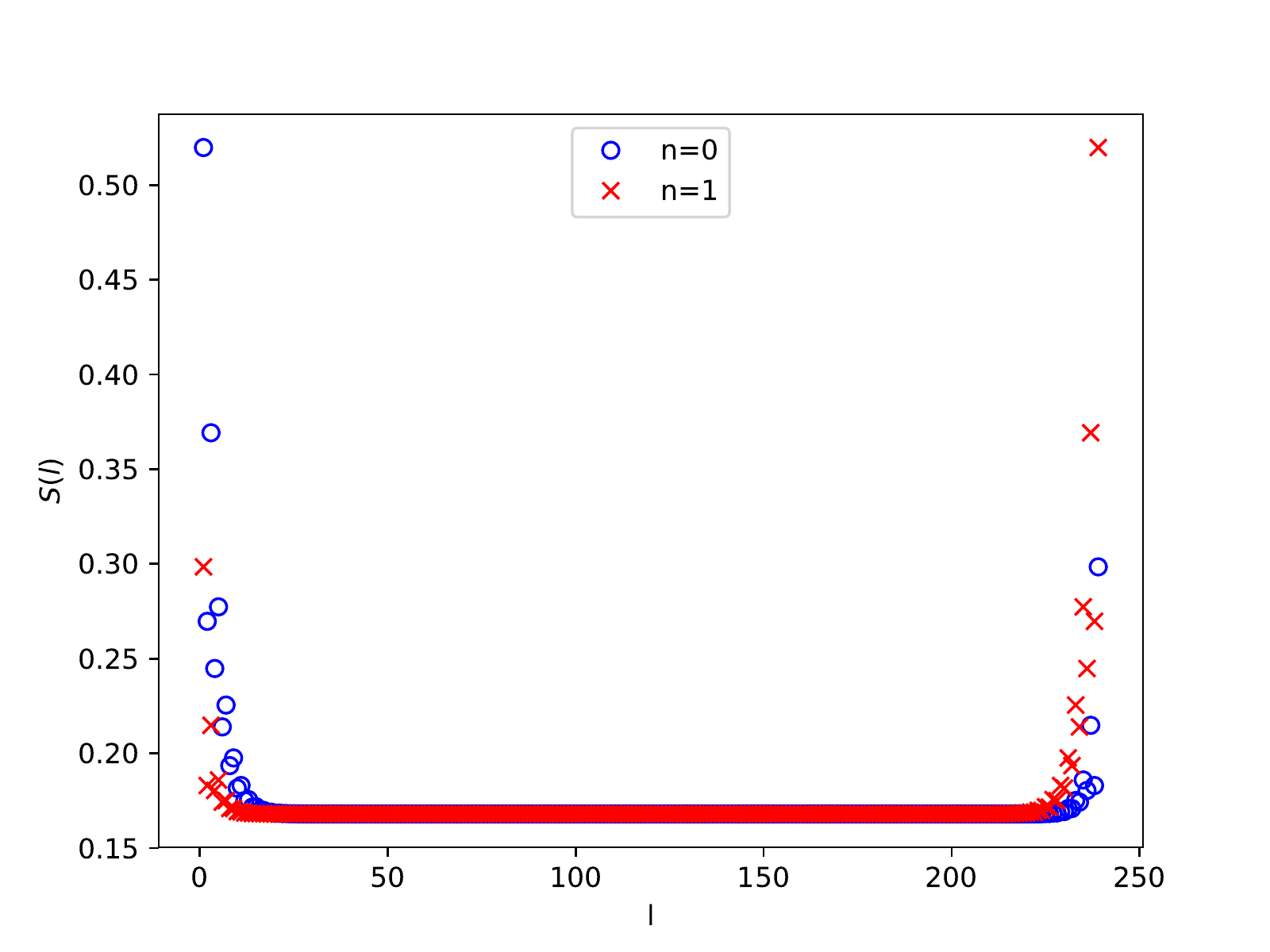}}

\subfloat[Occupation number.]{ \includegraphics[clip,width=\columnwidth]{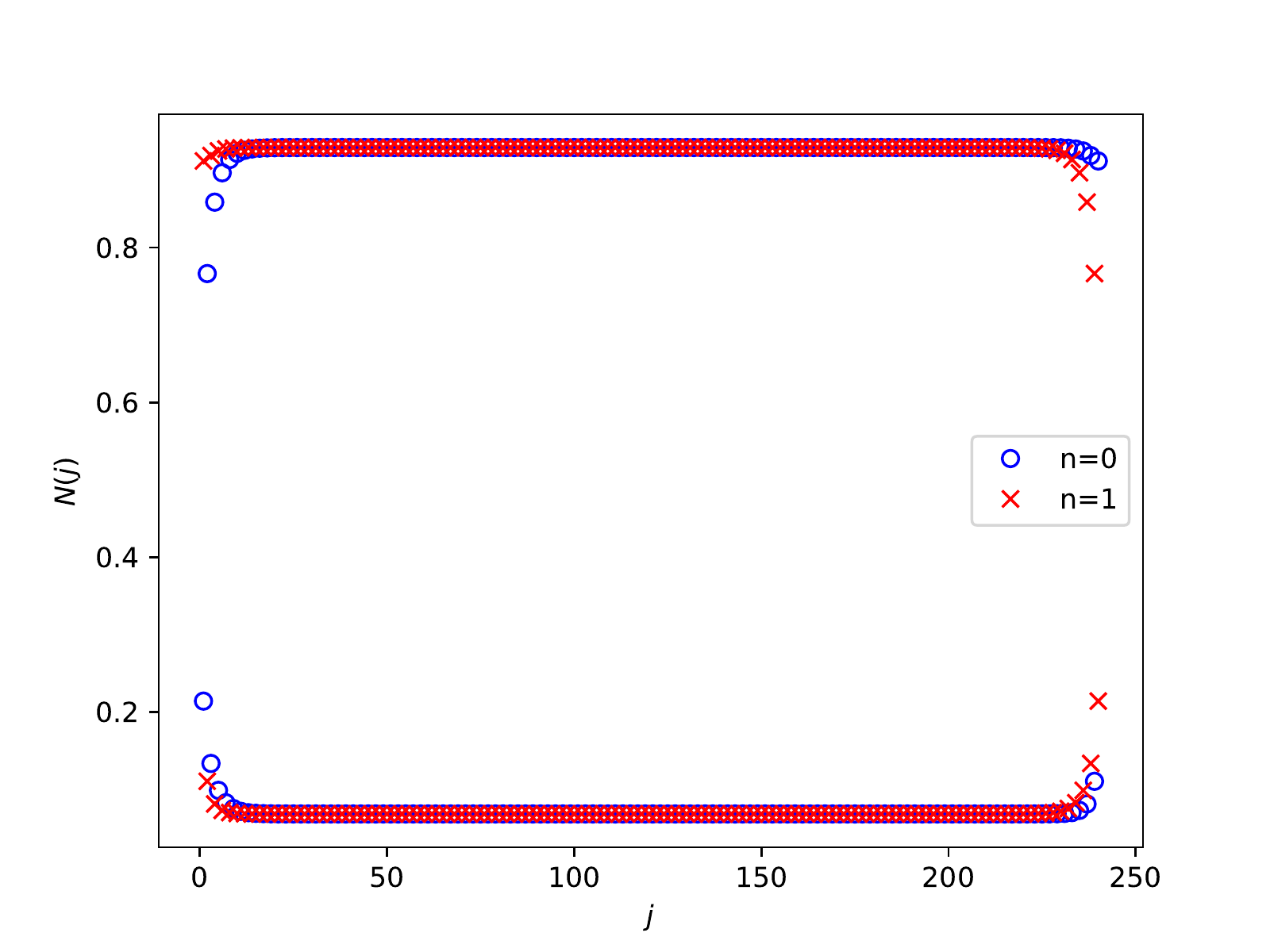}}
\caption{Observables for the ground state ($n=0$) and the first excited state ($n=1)$ at $h=0.7$ and $U=1.5$, a point in the CDW phase, are plotted. The system size is $L=240$.}
\label{fig-CDW_sample}
\end{figure}

By further increasing $U$, as we concluded above, we enter the regime where the ground state has CDW ordering. 
For $L=240$ and implementing the $h-$term as in Eq.~\eqref{eq-chemical-bonds}, the EE and occupation number for $h = 0.7$ and $U = 1.5$, are presented in Fig.~\ref{fig-CDW_sample}. In this case the two ground states belong to the even parity sector. In general, if we implement the $h-$term on bonds as we have in Eq.~\eqref{eq-chemical-bonds}, the ground state is doubly degenerate. For $L=4n$ (or $L = 4n+2$), where $n$ is a large integer, the two ground states belongs to the even (odd) sector. For $L$ odd, however, one ground state has even parity, the other odd. This can easily be understood in the limit where $U$ is infinite, so that one ignore the hopping and pairing terms in the Hamiltonian (which is then diagonal).
As is shown in Fig.~\ref{fig-CDW_sample} the EE for both ground states is saturated to a low value and the occupation number shows the pattern which one expects in the CDW phase, i.e. $\left( 1010 \cdots \right)$ and $\left( 0101\cdots \right)$.

If one instead implements the magnetic field on the sites, as in Eq.~\eqref{eq-chemical-sites}, the ground state for an even number of sites is still two-fold degenerate. However, for an odd number of sites, the ground state is unique, and the first excited state has energy $2 |h|$. Both of these states have the usual characteristics of CDW states.

\section{on the excited states and finite temperature properties}
\label{sec-excited_and_finite_T}
In this section, we study some properties of the model which are related to the full many-body spectrum of the model and its properties at finite temperature.

First of all we address the nature of the zero mode in the topological phase of the model. Usually in the studies on the topological phases the emphasis is on the degeneracy of the ground state. There are, however, a few cases where one can say more. For example in the Kitaev chain\cite{kitaev01} which can be written in terms of free fermions, the presence of the zero mode guarantees that the full many-body spectrum is doubly degenerate\cite{paul12}. Given an eigenstate in the many-body spectrum of the Kitaev chain with a given fermionic parity, one can construct another eigenstate with opposite parity. This can be done by looking at the occupancy of the zero mode. If it is empty, one can consider a state with an occupied zero mode and vice versa. Since filling the zero mode does not cost any energy (or an energy that is exponentially small in system size), the states are degenerate but with opposite parity. Therefore full many-body spectrum is doubly degenerate and the zero mode is called a \textit{strong} zero mode. Another, more non-trivial, example is the $XYZ$ chain for which Fendley found an operator which commutes with the Hamiltonian up to exponentially small corrections in the system size\cite{paul16}. The operator, however, is only normalizable in the topological phase, and hence the model hosts a strong zero mode and full many-body spectrum degeneracy in this phase. 

There is also a conjecture for general interacting models\cite{kells15} which states that if one starts from a non-interacting model within a topological phase, where we know there is a strong zero mode, one can construct a strong zero mode for the general interacting model providing that the system does not go through a phase transition or has level crossings. Therefore the full many-body spectrum would be degenerate all over the topological phase and there is equivalence between different parity sectors. In another study on non-integrable interacting models such as the Kitaev-Hubbard chain\cite{paul17}, 
it was argued that a strong zero mode is not present. Below, we present an analytic argument that
excludes the presence of a strong zero mode for a large part of the topological phase. We also study the edge magnetization in the region where our argument does not apply.

\subsection{Absence of strong zero mode in some regions within the topological phase} 

As we mentioned previously the Hamiltonian in Eq.~\eqref{eq-q-int-ising} is solvable for $U=0$ \cite{pfeuty70} and $h=0$ \cite{lsm61}, becuase it can be reduced to a free fermionic model in those cases. Based on the exact results we know that there is a strong zero mode all along the $U-$axis, except for the critical points $U=\pm 1$ and along the $h-$axis for $|h| < 1$. Thus it is natural to ask whether a strong zero mode survives throughout the topological phase. Here we present an argument to exclude the possibility of having a strong zero mode for some regions of the topological phase. 

\begin{figure}[h]
\includegraphics[width=\columnwidth]{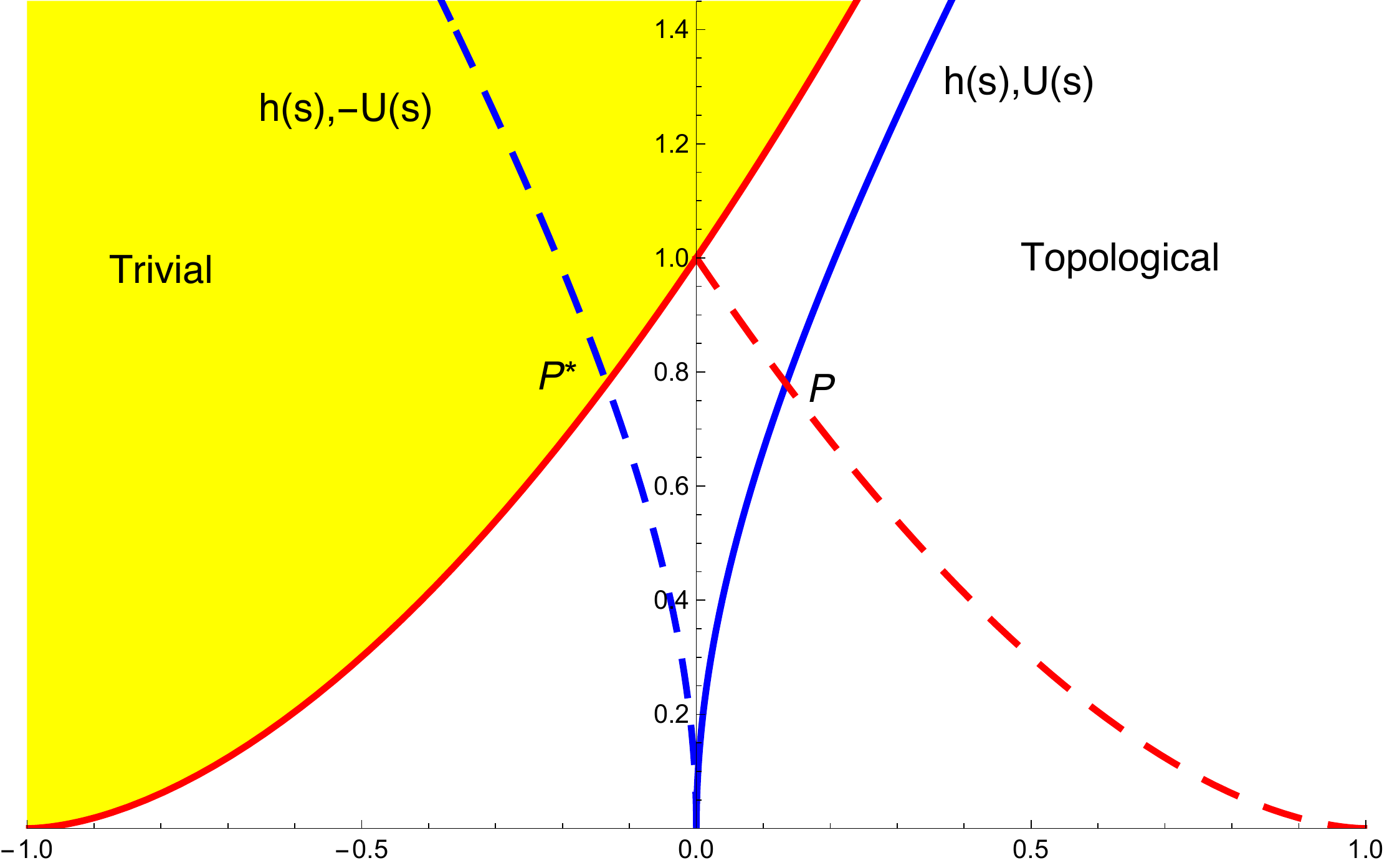}
\caption{(Color online) Part of the phase diagram of the model in Eq.~\eqref{eq-q-int-ising}. The solid red line separates the trivial and topological phase. The solid blue line parametrized by $s$, which lies in the topological phase, is an arbitrary line for which $H(h(s),U(s))$ has a doubly degenerate ground state. The dashed red and blue lines are the mirrors of the solid red and blue lines with respect to the $h-$axis. } 
\label{fig-strong_ZM_arg}
\end{figure}

Consider the solid blue line, $\left( h(s) , U(s) \right)$, in Fig.~\ref{fig-strong_ZM_arg}, which lies in the topological phase. Hence the Hamiltonian, $H(s)=H\left( h(s) , U(s) \right)$ has a doubly degenerate ground state. Actually, the line we chose is the Peschel-Emery line\cite{pe81},
for which the ground state
degeneracy is exact, the splitting is not exponentially small in system size.
We want to prove that the full many-body spectrum of $H(s)$ is \textit{not} doubly degenerate all along this line. To do so we examine the behaviour of the highest excited state(s). This can be done by noting that for any Hamiltonian $H$, the ground state(s) of $-H$ are the highest excited state(s) of $H$ itself. Let us consider $-H(s)$,
\begin{equation}
- H(h,U)= \sum_{j=1}^{L-1} \sigma^x_j\sigma^x_{j+1} + h \sum_{j=1}^{L} \sigma^z_j - U\sum_{j=1}^{L-1} \sigma^z_j\sigma^z_{j+1} \ .
\end{equation}
Now we perform an on-site rotation,
\begin{equation}
\sigma^x_j \rightarrow (-1)^j\sigma^x_j  \ , \hspace{0.5cm}\sigma^z_j \rightarrow - \sigma^z_j \ , 
\end{equation}
with which we get, 
\begin{align}
\label{eq-mirroe_H}
- H(h,U)&= - \sum_{j=1}^{L-1} \sigma^x_j\sigma^x_{j+1} - h \sum_{j=1}^{L} \sigma^z_j - U\sum_{j=1}^{L-1} \sigma^z_j\sigma^z_{j+1} \nonumber \\
& = H(h , -U) \ .
\end{align}

Therefore we can track the behaviour of the highest excited state(s) of $H(s)$, by looking at the ground state(s) of $H\left( h(s) , - U(s) \right)$. The dashed blue line in Fig.~\ref{fig-strong_ZM_arg} is the mirror of the solid blue line with respect to the $h-$axis. Following this dashed line from $h=0$ and $U=0$ till $P^*$, the ground state of $H$ is doubly degenerate. By passing $P^*$, however, we enter the trivial phase and hence the ground state becomes unique. We can use this information and interpret as the behaviour of the highest excited state(s) of $H(s)$, namely the Hamiltonian along the solid blue line. Doing so, we learn that by passing the point $P$, the highest excited state is \textit{not} doubly degenerate anymore. As a result there can not be a \textit{strong} zero mode and hence no equivalence of the parity sectors beyond the point $P$ along the blue solid line. Having this argument, we can draw the red dashed line which gives a bound on the region where the system potentially hosts a strong zero mode.

Below, we present numerical results which show that in part of this region, the edge magnetization survives for a long time, both for low as well as high temperature, which is consistent with the presence of a strong zero mode. We can not rigorously prove its existence, and it might well be that a strong zero mode is only present when the model reduces to a free-fermionic model.

\subsection{Time dependence of the edge magnetization}
In the following, we consider the time dependence of the edge magnetization\cite{paul17},
to see if the system has a strong zero mode. Long time coherence of edge magnetization
at infinite temperature is a signature of a strong zero mode\cite{paul17}. We will also consider
finite temperature, in which case the time-dependent edge magnetization at temperature $T$ is
given by
\begin{align}
A_{T} (t) =& \bigl< \sigma_1^x(t) \sigma_1^x (0) \bigr> \\
\nonumber
=&
\frac{1}{Z} \sum_{j_1,j_2}
e^{-\epsilon_{j_1}/(k T)} e^{i (\epsilon_{j_1}-\epsilon_{j_2})t}
| \bigl< j_2 \bigr | \sigma_1^x \bigl| j_1 \bigr> |^2  \ ,  
\end{align}
where $\epsilon_j$ are the energies of the system. We will refer to the time in units of $1/J$, and
temperature in units of $J$, where $J=1$. We considered system sizes up to $L=16$.
\begin{figure}[htp]
\subfloat[]{\includegraphics[clip,width=\columnwidth]{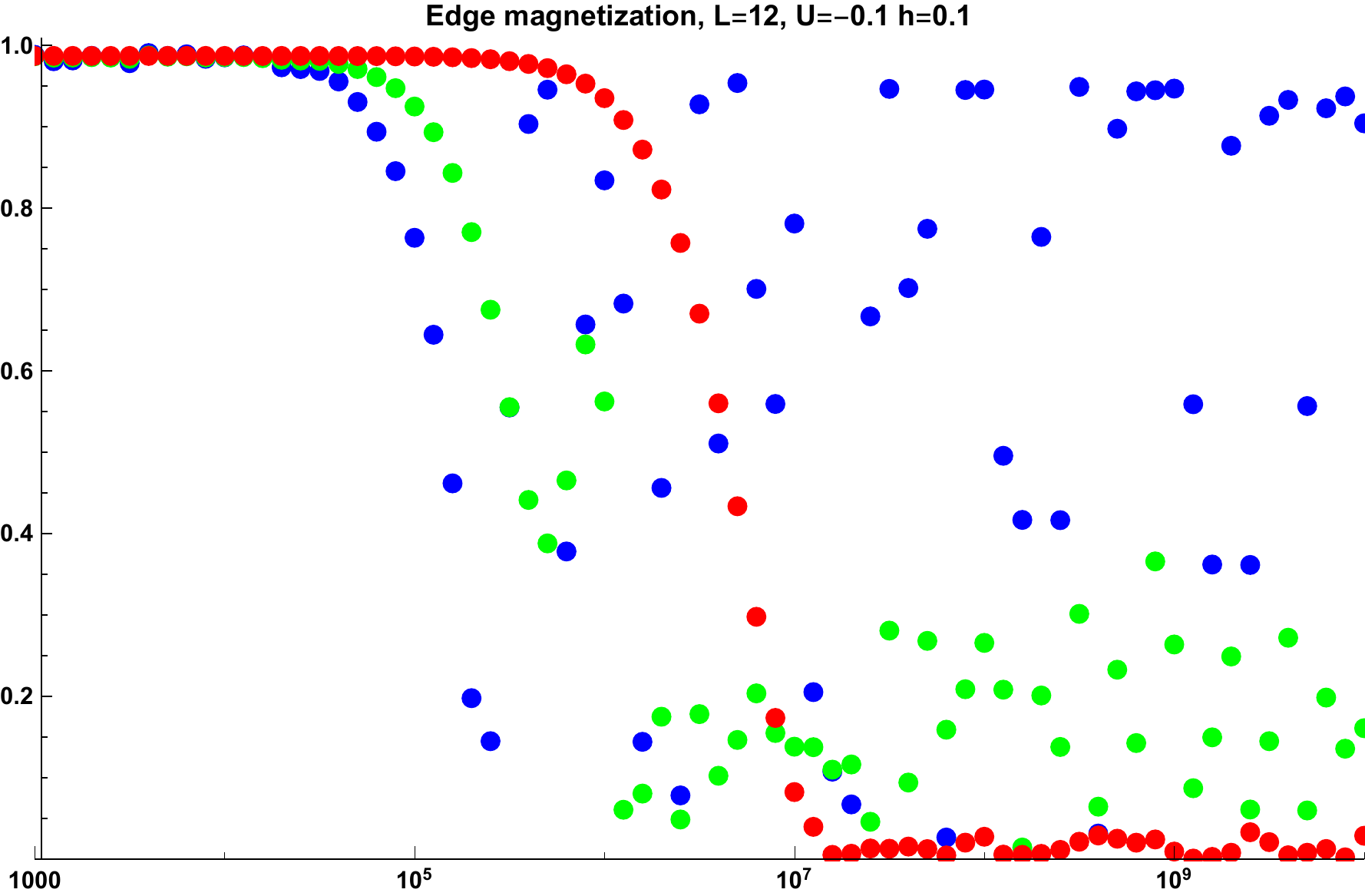}}

\subfloat[]{ \includegraphics[clip,width=\columnwidth]{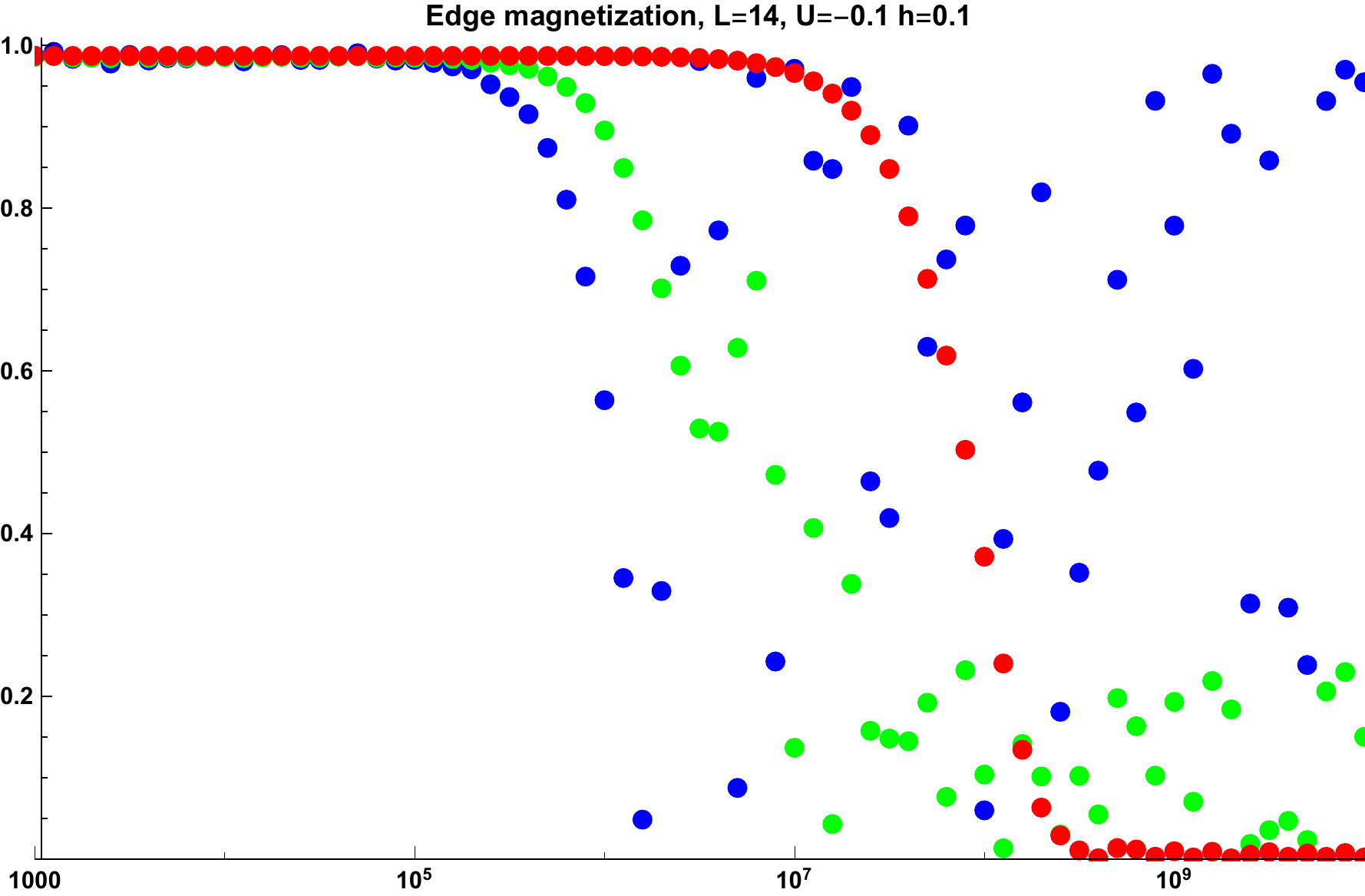}}

\subfloat[]{ \includegraphics[clip,width=\columnwidth]{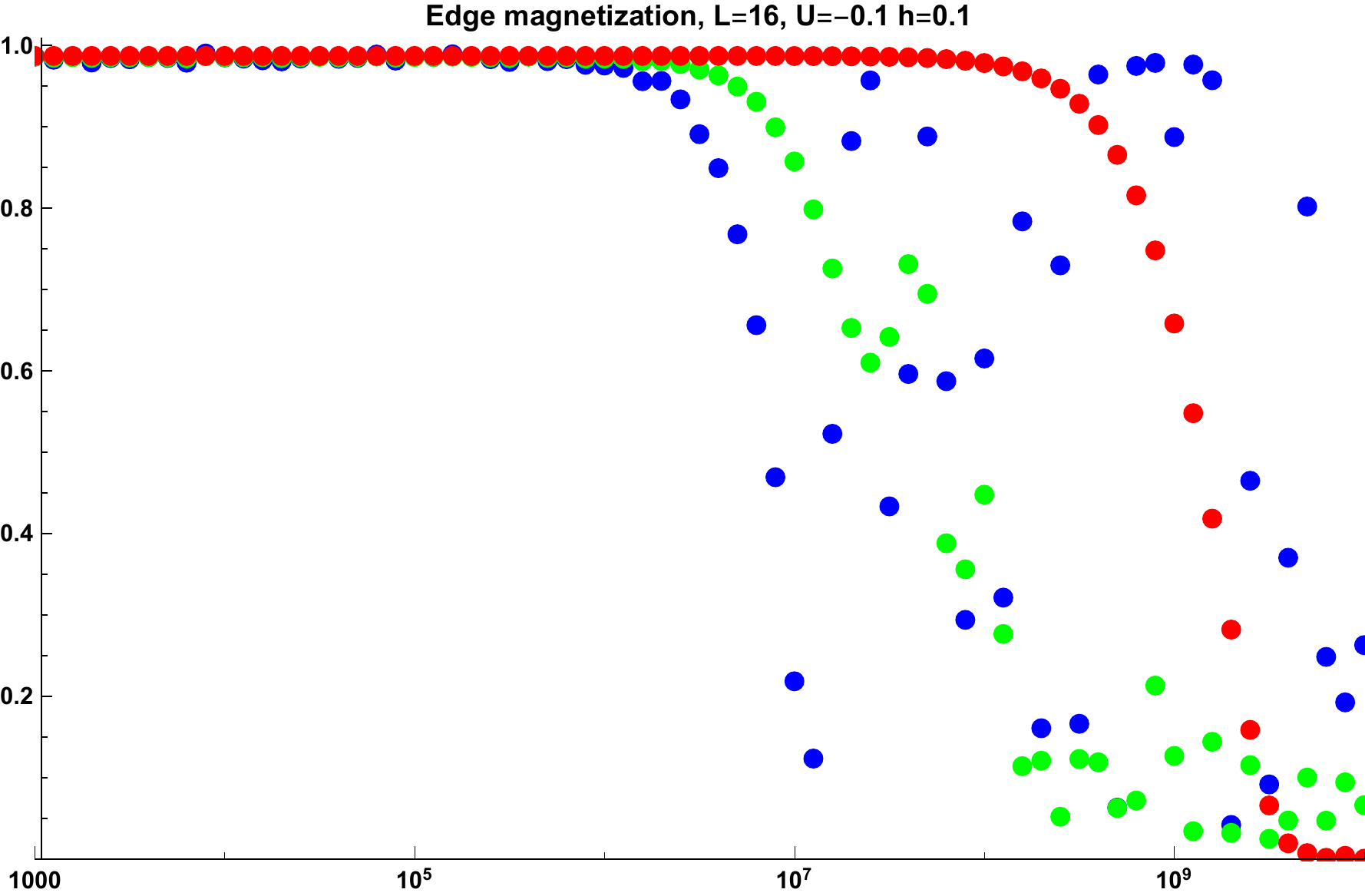}}
\caption{Edge magnetization $|A_T (t)|$ for $U = -0.1$, $h = 0.1$, as a function of time, for
$T=10^{-3},1,10^{3}$, blue, green, red symbols, for (a) $L=12$, (b) $L=14$ and (c) $L=16$. Note that the logarithmic time axis starts at $10^3$. }
\label{fig:U=-01h=01}
\end{figure}

We first look at small $h = 0.1$, and a rather strong interaction, $| U | = 0.5$, but still in the region in Fig.~\ref{fig-strong_ZM_arg} where the analytic argument can not exclude a strong
zero-mode. In these two cases, we find that the edge magnetization at very large $T=10^3$, so
effectively infinite temperature, survives up to $t\sim 100$, before
dropping to zero, with only little
variation upon increasing the system size from $L=10$ to $L=14$. For $T = 1$, one obtains roughly
the same result, while for $T = 10^{-3}$, the long-time edge magnetization oscillates around $0.7$,
with an amplitude of $0.1$. For these parameters, it is clear that the system does not have a strong
zero mode. 

For $h = 0.1$ and $| U | = 0.1$, the situation is rather different, as we show in Fig.~\ref{fig:U=-01h=01}
for $h = 0.1$ and $U = -0.1$, where we plot $|A_T (t) |$ as a function of $t$ on a log scale for
temperatures $T = 10^{-3},1,10^{3}$, with blue, green and red symbols respectively. The magnetic field
was implemented on the bonds, see Eq.~\eqref{eq-chemical-bonds}.

We observe that the edge magnetization survives longer upon increasing the system size,
for all temperatures.
Interestingly, the edge magnetization survives longer for higher temperatures.
However, at high temperature, it drops to zero, while for low temperature, the
edge magnetization rapidly oscillates at long times.
The high temperature behaviour is consistent with the behaviour seen for
the XYZ chain\cite{paul17}, for which it was proven that the system has a strong zero mode\cite{paul16}.
\begin{figure}[t]
\includegraphics[width=\columnwidth]{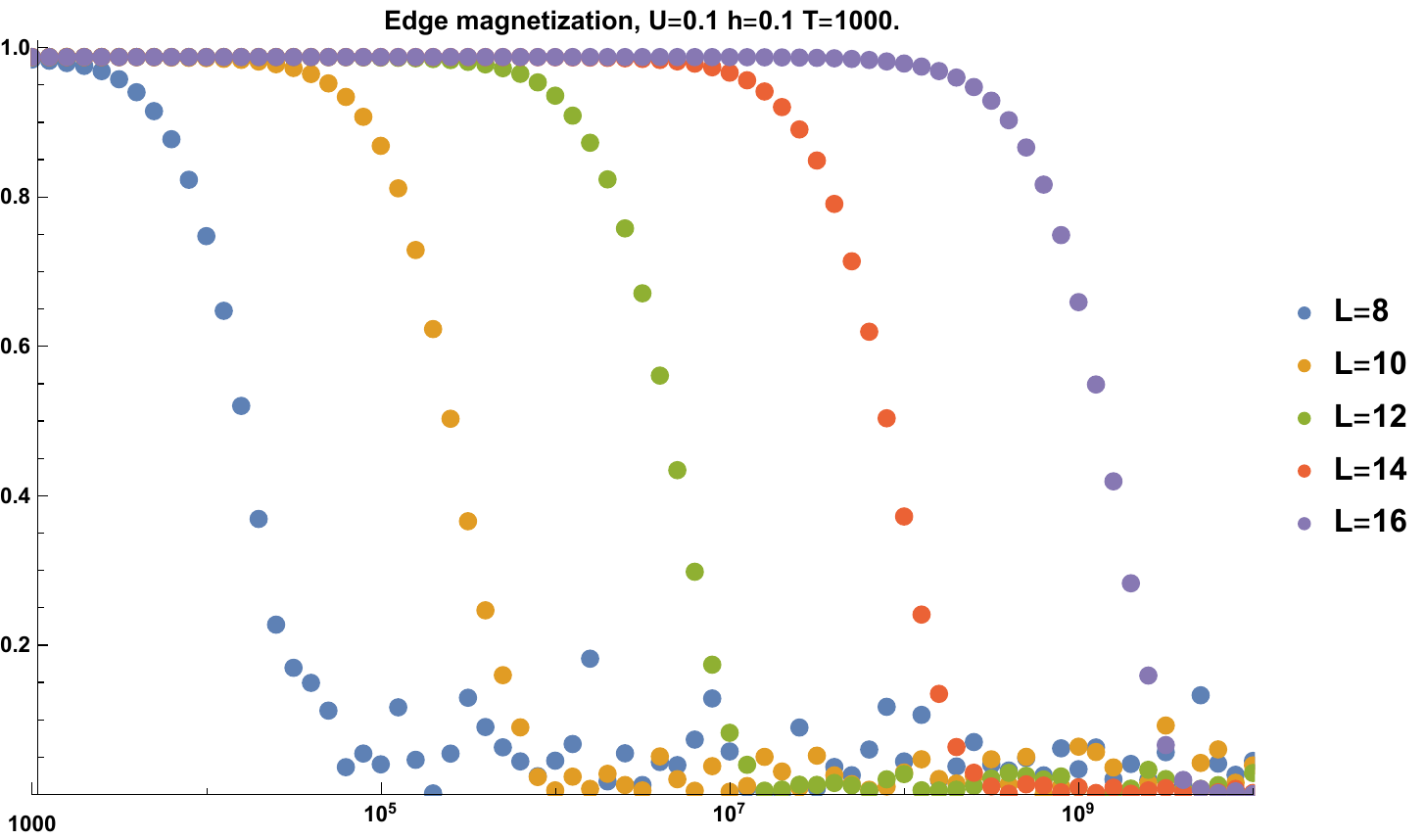}
\caption{The edge magnetization $|A_{T=1000} (t)|$ for $U=0.1$, $h=0.1$ and system sizes
$L = 8,10,12,14,16$ versus time on a logarithmic scale.}
\label{fig:em-variousL}
\end{figure}
In Fig.~\ref{fig:em-variousL}, we plot the $T=1000$ edge magnetization for $U=0.1$, $h=0.1$ and
even system sizes $L= 8 - 16$. We observe that the time that the edge magnetization survives grows exponentially
with system size, at least for the system sizes that we could consider.
So based on this data, we can not rule out the presence of a strong zero mode for these parameters.

We note that for $U<0$, the different ways
of implementing the magnetic field give very similar results. However, for $U>0$ at low temperature,
the time at which the edge magnetization drops to zero for the first time can differ by as much as two
orders of magnitude. For which implementation of the magnetic field this happens first, depends on
the system size. That the time dependence of the edge magnetization depends on the value of the
magnetic field on the edges is of course not so strange, but from the numerical results we obtained,
a clear picture does not emerge.

For larger magnetic fields and moderate interaction, $h = 0.7$, and $|U| = 0.1$, the edge magnetization
for infinite temperature survives to $t\sim 100$, again with little dependence on the system size. Hence,
for these parameters, there is clearly no strong zero mode. For $h=0.7$ and $U=0.5$, our
analytical argument excludes a strong zero mode, because the point $h=0.7$, $U=-0.5$ is in the trivial
phase. Indeed, we found that at high temperature the edge magnetization drops sharply at $t\sim 1$,
while the edge magnetization for low temperature, probing the ground state, survives up to long times.

\section{Discussion}
\label{sec-discussion}

We presented a detailed study the phase diagram of the Kitaev-Hubbard chain, that is the Kitaev chain in the presence of a nearest-neighbour density-density interaction. In the case of attractive interactions, the model exhibits a topological phase and a trivial phase, the same phases which also appear in the Kitaev chain. For repulsive interactions, however, the phase diagram is more interesting, with an IC phase, an esCDW and a CDW phase, in addition to the trivial and topological phases.

The bosonic incarnation of the interacting Kitaev chain is an interacting transverse field Ising chain, which is dual to the quantum ANNNI model. There was an open question in the quantum ANNNI model about the presence of the floating phase for high value of frustration. Using fermionic terminology, this translates to presence of the IC phase in the Kitaev-Hubbard chain for small values of the transverse field, $h$, that is, close to the $XY$ critical point. Using DMRG we see the presence of the IC down to $h=0.05$ which corresponds to $\kappa\simeq 20$, although it becomes quite narrow. Therefore we strongly believe that the IC phase continues to $h=0$, which is the limit of infinite frustration. 

The IC phase is a gapless phase, for which the EE can be fitted to the CC formula with a value of the central charge $c \simeq 1$. However, the EE as a function of sub-system size shows additional
oscillations, on top of the CC behaviour. It would be interesting to find a good ansatz for the oscillations, 
because a quantitative knowledge of these oscillations can shed additional light on the nature of the IC phase. It would also be interesting to see how well they correlate with the oscillations observed in
the site occupation number.  

The esCDW phase, we believe, was missed in previous studies. This phase only appears for an even number of sites and its ground state looks like an excited state in the CDW phase. In this region, the model is gapless, with a dynamical critical exponent $z \simeq 1.8$. The EE in this phase grows as a function of subsystem size, but can not be fitted to the CC formula, which is consistent with the dynamical critical exponent $z>1$. For odd system sizes, the model shows a CDW phase, with a unique ground state. It is quite intriguing that there is finite region in the phase diagram which responds to the (parity of) the number of sites. In infinite system size studies, one observes a normal CDW phase\cite{nagy}. This is consistent with our findings, because the ground state energy per site is slightly lower, of order $10^{-3}$, for odd system sizes. Because the energy difference is so small, the nature of this phase will in practice be determined by other perturbations, for instance additional interaction terms. 

For repulsive interactions, we found that there are four phases which emerge from the $XY$ critical point, namely the topological phase, the IC phase, the esCDW phase and the CDW phase. Capturing these phases and finding the relevant operator(s) for each phase using for instance bosonization will be left for future research.

Finally, we considered the nature of the zero mode in the topological phase. We presented an analytical argument that excludes a strong zero mode for a large part of the topological phase. To address the remaining region, we calculated the time dependence of the edge magnetization\cite{paul17}. In the remaining region, these results rule out a strong zero mode in the case that either $h$ or $|U|$ is large. For small $h$ and $|U|$, we can not exclude a strong zero mode. For $U>0$, the time dependence of the edge magnetization does not show consistent behaviour as a function of temperature and the implementation of the magnetic field. This warrants further study of the edge magnetization in this model.

{\em Acknowledgments ---}
We would like to thank P.~Fendley, M.~Hermanns, L.~Herviou, A.~Kantian, G.~Kells and D.~Schuricht
for interesting discussions. This work was sponsored, in part, by the Swedish Research Council.


\begin{thebibliography}{99}

\bibitem{class08}
A.P.~Schnyder, S.~Ryu, A.~Furusaki, A.W.W.~Ludwig,
{\it Classification of topological insulators and superconductors in three spatial dimensions},
Phys. Rev. B. {\bf 78}, 195125 (2008),
\doi{10.1103/PhysRevB.78.195125}.

\bibitem{class10}
S.~Ryu, A.P.~Schnyder, A.~Furusaki, A.W.W.~Ludwig,
{\it Topological insulators and superconductors: tenfold way and dimensional hierarchy},
New J. Phys. {\bf 12}, 065010 (2010),
\doi{10.1088/1367-2630/12/6/065010}.

\bibitem{class-RMP}
C.-K.~Chiu, J.C.Y.~Teo, A.P.~Schnyder,S.~Ryu,
{\it Classification of topological quantum matter with symmetries},
Rev. Mod. Phys. {\bf 88}, 035005 (2016),
\doi{10.1103/RevModPhys.88.035005}. 


\bibitem{kitaev09}
A.Y.~Kitaev, 
{\it Periodic table for topological insulators and superconductors},
AIP Conf. Proc. {\bf 1134} 22 (2009),
\doi{10.1063/1.3149495}. 

\bibitem{kitaev01}
A.Y.~Kitaev,
{\it Unpaired Majorana fermions in quantum wires},
Phys. Usp. {\bf 44}, 131 (2001),
\doi{10.1070/1063-7869/44/10S/S29}.

\bibitem{fk10}
L.~Fidkowski, A.~Kitaev,
{\it Effects of interactions on the topological classification of free fermion systems},
Phys. Rev. B. {\bf 81}, 134509 (2010),
\doi{10.1103/PhysRevB.81.134509}. 

\bibitem{fk11}
L.~Fidkowski, A.~Kitaev,
{\it Topological phases of fermions in one dimension},
Phys. Rev. B. {\bf 83}, 075103 (2011),
\doi{10.1103/PhysRevB.83.075103}. 


\bibitem{alicea-fisher11}
E.M.~Stoudenmire, J.~Alicea, O.A.~Starykh, M.P.A. Fisher,
{\it Interaction effects in topological superconducting wires supporting Majorana fermions}
Phys. Rev. B. {\bf 84}, 014503 (2011),
\doi{10.1103/PhysRevB.84.014503}. 

\bibitem{selke}
W.~Selke,
{\it The ANNNI model -- Theoretical analysis and experimental application},
Phys. Rep. {\bf 170}, 213 (1988),
\doi{10.1016/0370-1573(88)90140-8}.

\bibitem{feo06}
M.~Beccaria, M.~Campostrini, A.~Feo,
{\it Density-matrix renormalization-group study of the disorder line in the quantum axial next-nearest-neighbor Ising model},
Phys. Rev. B. {\bf 73}, 052402 (2006),
\doi{10.1103/PhysRevB.73.052402}.


\bibitem{feo07}
M.~Beccaria, M.~Campostrini, A.~Feo, 
{\it Evidence for a floating phase of the transverse ANNNI model at high frustration}, 
Phys. Rev. B. {\bf76}, 094410 (2007), 
\doi{10.1103/PhysRevB.76.094410}.

\bibitem{nagy}
A.~Nagy, 
{\it Exploring phase transitions by finite-entanglement scaling of MPS in the 1D ANNNI model}, 
New. J. Phys. {\bf 13}, 023015 (2011), 
\doi{10.1088/1367-2630/13/2/023015}.

\bibitem{sela11}
E.~Sela, R.G.~Pereira,
{\it Orbital multicriticality in spin-gapped quasi-one-dimensional antiferromagnets},
Phys. Rev. E. {\bf 84}, 014407 (2011),
\doi{10.1103/PhysRevB.84.014407}.

\bibitem{allen}
D.~Allen, P.~Azaria, P.~Lecheminant,
{\it A two-leg quantum Ising ladder: a bosonization study of the ANNNI model},
J. Phys. A {\bf 34}, L305 (2001),
\doi{10.1088/0305-4470/34/21/101}.


\bibitem{white92}
S.R.~White,
{\it Density matrix formulation for quantum renormalization groups},
Phys. Rev. Lett. {\bf 69}, 2863 (1992),
\doi{10.1103/PhysRevLett.69.2863}.

 
\bibitem{scholl05}
U.~Schollw\"ock,
{\it The density-matrix renormalization group},
Rev. Mod. Phys. {\bf 77}, 259 (2005),
\doi{10.1103/RevModPhys.77.259}.

\bibitem{onsager}
L.~Onsager, {\it Crystal statistics. I. A two-dimensional model with an order-disorder transition},
Phys. Rev. {\bf 65}, 117 (1944),
\doi{10.1103/PhysRev.65.117}.


\bibitem{baxter}
R.J.~Baxter, {\it Exactly solved models in statistical mechanics},
Academic Press (1982).

\bibitem{kardar}
M.~Kardar, {\it Statistical physics of fields},
Cambridge University Press (2007).

\bibitem{goldenfeld}
N.~Goldenfeld, {\it Lectures on phase transitions and the renormalization group},
Addison Wesley(1992).

\bibitem{sf80}
W.~Selke, M.E.~Fisher,
{\it Two-Dimensional Ising Models with Competing Interactions -
A Monte Carlo Study},
Z. Physik B {\bf 40}, 71(1980),
\doi{10.1007/BF01295073}.

\bibitem{selke81}
W.~Selke,
{\it Finite-Size Behaviour of the Two-Dimensional ANNNI Model},
Z. Physik B {\bf 43}, 335 (1981),
\doi{10.1007/BF01292801}.

\bibitem{epw70}
R.J.~Elliott, P.~Pfeuty, C.~Wood,
{\it Ising Model with a Transverse Field},
Phys. Rev. Lett. {\bf 25}, 443 (1970),
\doi{10.1103/PhysRevLett.25.443}.


\bibitem{pfeuty70}
P.~Pfeuty,
{\it The one-dimensional Ising model with a transverse field},
Ann. Phys. {\bf 57}, 79 (1970),
\doi{10.1016/0003-4916(70)90270-8}.

\bibitem{fs78}
E.~Fradkin, L.~Susskind,
{\it Order and disorder in gauge systems and magnets},
Phys. Rev. D {\bf 17}, 2637 (1978).


\bibitem{jw28}
P.~Jordan, E.~Wigner,
{\it \"Uber das Paulische \"Aquivalenzverbot},
Z. Physik {\bf 47}, 631 (1928),
\doi{10.1007/BF01331938}.

\bibitem{lsm61}
E.~Lieb, T.~Schultz, D.~Mattis,
{\it Two soluble models of an antiferromagnetic chain},
Ann. Phys. {\bf 16}, 407 (1961),
\doi{10.1016/0003-4916(61)90115-4}.



\bibitem{alt11}
E.~Sela, A.~Altland, A.~Rosch,
{\it Majorana fermions in strongly interacting helical liquids},
Phys. Rev. B {\bf 84}, 085114 (2011),
\doi{10.1103/PhysRevB.84.085114}.

\bibitem{hs12}
F,~Hassler, D.~Schuricht,
{\it Strongly interacting Majorana modes in an array of
Josephson junctions},
New J. Phys. {\bf 14}, 125018 (2012),
\doi{10.1088/1367-2630/14/12/125018}. 


\bibitem{pe81}
I.~Peschel, V.J.~Emery,
{\it Calculation of spin correlations in two-dimensional Ising systems from one-dimensional kinetic models},
Z. Phys. B {\bf 43}, 241 (1981),
\doi{10.1007/BF01297524}.

\bibitem{katsura15}
H.~Katsura, D.~Schuricht, M.~Takahashi,
{\it Exact ground states and topological order in interacting Kitaev/Majorana chains},
Phys. Rev. B {\bf 92}, 115137 (2015),
\doi{10.1103/PhysRevB.92.115137}.

\bibitem{ma18}
I.~Mahyaeh, E.~Ardonne,
{\it Exact results for a $\mathcal{Z}_3$-clock-type model and some close relatives},
Phys. Rev. B {\bf 98}, 245104 (2018),
\doi{10.1103/PhysRevB.98.245104}. 


\bibitem{gergs}
N.M.~Gergs, L.~Fritz, D.~Schuricht,
{\it Topological order in the Kitaev/Majorana chain in the presence of disorder and interactions},
Phys. Rev. B {\bf 93}, 075129 (2016),
\doi{10.1103/PhysRevB.93.075129}. 


\bibitem{book:shankar}
R.~Shankar,
{\it Quantum field theory and condensed matter - An introduction},
Cambridge University press, 2017,
\doi{10.1017/9781139044349}.


\bibitem{alps07}
F.~Albuquerque, F.~Alet, P.~Corboz, P.~Dayal, A.~Feiguin, S.~Fuchs, L.~Gamper, E.~Gull, S.~G\"urtler, A.~Honecker, R.~Igarashi, M.~K\"orner, A.~Kozhevnikov, A.~L\"auchli, S.R.~Manmana, M.~Matsumoto, I.P.~McCulloch, F.~Michel, R.M.~Noack, G.~Pawlowski, L.~Pollet, T.~Pruschke, U.~Schollw\"ock, S.~Todo, S.~Trebst, M.~Troyer, P.~Werner, S.~Wessel, and for the ALPS collaboration,
{\it The ALPS project release 1.3: Open-source software for strongly correlated systems},
 J. Magn. Magn. Mater. {\bf 310}, 1187 (2007),
 \doi{10.1016/j.jmmm.2006.10.304}.

\bibitem{alps11}
 B.~Bauer, L.D.~Carr, H.G.~Evertz, A.~Feiguin, J.~Freire, S.~Fuchs, L.~Gamper, J.~Gukelberger, E.~Gull, S. G\"uertler, A.~Hehn, R.~Igarashi, S.V.~Isakov, D.~Koop, P.N.~Ma, P.~Mates, H.~Matsuo, O.~Parcollet, G.~Pawlowski, J.D.~Picon, L.~Pollet, E.~Santos, V.W.~Scarola, U.~Schollw\"ock, C.~Silva, B.~Surer, S.~Todo, S.~Trebst, M.~Troyer, M.L.~Wall, P.~Werner, S.~Wessel,
 {\it The ALPS project release 2.0: open source software for strongly correlated systems},
 J. Stat. Mech (2011) P05001,
 \doi{10.1088/1742-5468/2011/05/P05001}.
 
 \bibitem{alps14}
 M.~Dolfi, B.~Bauer, S.~Keller, A.~Kosenkov, T.~Ewart, A.~Kantian, 
 T.~Giamarchi, M.~Troyer, 
 {\it Matrix product state applications for the ALPS project}, 
 Comput. Phys. Commun. {\bf 185}, 3430 (2014),
 \doi{10.1016/j.cpc.2014.08.019}. 
 
 \bibitem{fb72}
 M.E.~Fisher, M.N.~Barber,
{\it  Scaling Theory for Finite-Size Effects in the Critical Region},
Phys. Rev. Lett. {\bf 28}, 1516(1972),
\doi{10.1103/PhysRevLett.28.1516}.
 
\bibitem{sachdevbook}
S.~Sachdev, {\it Quantum phase transitions},
Cambridge University Press (2011).

\bibitem{BYB}
P.~ Di Francesco , P.~ Mathieu, D.~S\'{e}n\'{e}chal, 
{\it Conformal Field Theory },
New York: Springer (1999).


\bibitem{cc04}
P.~Calabrese, J.~Cardy, 
{\it Entanglement entropy and quantum field theory}, 
J. Stat. Mech.(2004) P06002,
\doi{10.1088/1742-5468/2004/06/P06002}.

\bibitem{cc09}
P.~Calabrese, J.~Cardy, 
{\it Entanglement entropy and conformal field theory}, 
J. Phys. A: Math. Theor. {\bf 42}, 504005 (2009),
\doi{10.1088/1751-8113/42/50/504005}.

\bibitem{wilczek94}
C.~Holzhey, F.~Larsen, F.~Wilczek,
{\it Geometric and renormalized entropy in conformal field theory},
Nuc. Phys. B {\bf 424}, 443 (1994),
\doi{10.1016/0550-3213(94)90402-2}. 

\bibitem{hughes15}
Y.~Zhuang, H.J.~Changlani, N.M.~Tubman,T.L.~Hughes,
{\it Phase diagram of the Z3 parafermionic chain with chiral interactions},
Phys. Rev. B. {\bf 92}, 035154(2015),
\doi{10.1103/PhysRevB.92.035154}. 


\bibitem{paul12}
P.~Fendley,
{\it Parafermionic edge zero modes in $Z_n$-invariant spin chains},
J. Stat. Mech. P11020 (2012),
\doi{10.1088/1742-5468/2012/11/P11020}.

\bibitem{paul16}
P.~Fendley,
{\it Strong zero modes and eigenstate phase transitions in the XYZ/interacting Majorana chain}, 
J. Phys. A: Math. Theor. {\bf 49}, 30LT01 (2016),
\doi{10.1088/1751-8113/49/30/30LT01}.

\bibitem{kells15}
G.~Kells,
{\it Many-body Majorana operators and the equivalence of parity sectors},
Phys. Reb. B {\bf 92}, 081401(R) (2015),
\doi{10.1103/PhysRevB.92.081401}. 

\bibitem{paul17}
J.~Kemp, N.Y.~Yao, C.R.~Laumann, P.~Fendley,
{\it Long coherence times for edge spins},
J. Stat. Mech. 063105 (2017),
\doi{10.1088/1742-5468/aa73f0}.



\end{thebibliography}
\end{document}